\documentclass[aps,nofootinbib,preprintnumbers]{revtex4}
\usepackage{eurosym}
\usepackage{CJK}
\usepackage{amsfonts}
\usepackage{amsmath}
\usepackage{amssymb}
\usepackage[english]{babel}
\usepackage{graphicx}
\usepackage{epsfig}
\usepackage{bm}
\usepackage{verbatim}
\usepackage[utf8]{inputenc}
\usepackage{booktabs}
\usepackage{multirow}
\usepackage{braket}
\usepackage{slashed}
\usepackage{xcolor}
\usepackage[colorlinks=true,urlcolor=red,citecolor=red]{hyperref}
\usepackage{float}
\usepackage{blindtext}
\usepackage{placeins}
\usepackage[utf8]{inputenc}
\usepackage{bbm}
\usepackage[colorinlistoftodos]{todonotes}
\usepackage{mdframed}
\usepackage{tabularx}
\newcolumntype{L}{>{\raggedright\arraybackslash}X}
\usepackage{subfigure}

\input{tcilatex}
\begin{document}

\title{Two-loop neutrino mass model with modular $S_4$ symmetry.}
\author{A. E. C\'arcamo Hern\'andez}
\email{antonio.carcamo@usm.cl}
\affiliation{{Universidad T\'ecnica Federico Santa Mar\'{\i}a, Casilla 110-V, Valpara%
\'{\i}so, Chile}}
\affiliation{{Centro Cient\'{\i}fico-Tecnol\'ogico de Valpara\'{\i}so, Casilla 110-V,
Valpara\'{\i}so, Chile}}
\affiliation{{Millennium Institute for Subatomic Physics at High-Energy Frontier
(SAPHIR), Fern\'andez Concha 700, Santiago, Chile}}
\author{J. Echeverria-Puentes}
\email{j.echeverriapuentes@uandresbello.edu}
\affiliation{Universidad T\'{e}cnica Federico Santa Mar\'{\i}a, Casilla 110-V, Valpara%
\'{\i}so, Chile.}
\affiliation{{Centro Cient\'{\i}fico-Tecnol\'ogico de Valpara\'{\i}so, Casilla 110-V,
Valpara\'{\i}so, Chile}}
\affiliation{Departamento de Ciencias F\'isicas, Universidad Andres Bello, \\
Sazi\'e 2212, Piso 7, Santiago, Chile}
\author{Vishnudath K. N.}
\email{vishnudath.kn@vit.ac.in}
\affiliation{Department of Physics, School of Advanced Sciences, Vellore Institute of Technology,
Vellore 632014, Tamilnadu, India}
\author{Sergey Kovalenko}
\email{sergey.kovalenko@unab.cl}
\affiliation{Departamento de Ciencias F\'isicas, Universidad Andres Bello, \\
Sazi\'e 2212, Piso 7, Santiago, Chile}
\affiliation{Millennium Institute for Subatomic Physics at High-Energy Frontier
(SAPHIR), Fern\'andez Concha 700, Santiago, Chile}
\affiliation{Dzhelepov Laboratory of Nuclear Problems, JINR,
 141980 Dubna, Russia}

\author{Daniel Salinas-Arizmendi}
\email{daniel.salinas@usm.cl}
\affiliation{Departamento de F\'isica, Universidad T\'{e}cnica Federico Santa Mar\'{\i}a,
Casilla 110-V, Valpara\'{\i}so, Chile}
\affiliation{Centro Cient\'{\i}fico-Tecnol\'{o}gico de Valpara\'{\i}so, Casilla 110-V,
Valpara\'{\i}so, Chile}
\author{Carlos A. Vaquera-Araujo}
\email{vaquera@fisica.ugto.mx}
\affiliation{Secretar\'ia de Ciencia, de Humanidades, Tecnolog\'ia e Innovaci\'on, Insurgentes Sur 1582. Colonia Cr\'edito Constructor, Benito Ju\'arez 03940, Ciudad de M\'exico, Mexico}
\affiliation{Departamento de F\'isica, Divisi\'on de Ciencdias e Ingenier\'ias, Campus Le\'on, Universidad de
  Guanajuato, Loma del Bosque 103, Lomas del Campestre, Le\'on 37150, Guanajuato, Mexico}
\affiliation{Dual CP Institute of High Energy Physics, Colima 28045, Colima, Mexico}

\begin{abstract}
We propose a two-loop radiative neutrino mass model based on the modular $\Gamma_4 \simeq S_4$ flavour symmetry supplemented by a discrete $Z_3$ symmetry. After spontaneous modular symmetry breaking, a remnant $Z_2$ symmetry  guarantees both the radiative origin of active neutrino masses and stabilizes dark matter candidates. 
The model successfully reproduces charged lepton masses and neutrino oscillation data for normal ordering. It also predicts observable rates for charged lepton flavour violation (LFV). Due to the singlet-doublet mixing the model provides a viable scalar dark matter candidate. A fermionic dark matter candidate, strongly linked to LFV, is also present.
We identify parameter regions consistent with relic density, LFV constraints, and direct detection limits, providing testable benchmark configurations.
\end{abstract}

\author{}
\maketitle
%%%%%%%%%%%%%%%%%%%%

\section{Introduction}

The Standard Model (SM) provides a successful description of fundamental particle interactions but leaves some fundamental questions unanswered. Among the most compelling ones are the mechanism that generates neutrino masses, the nature of dark matter (DM), and the underlying structure responsible for the observed hierarchy of lepton masses and mixings. These open problems strongly motivate the exploration of extensions of the SM where neutrino physics, flavour structure, and dark sector phenomenology can be addressed within a common framework.

Radiative neutrino mass models, in particular scotogenic models, are especially appealing in this context, since they can relate the smallness of neutrino masses to loop-suppressed mechanisms while naturally introducing new particles at experimentally accessible scales, see for instance Ref.~\cite{Cai:2017jrq} for a review and Refs.~\cite{Jana:2019mgj,Arbelaez:2022ejo} for comprehensive studies of one- and two-loop radiative neutrino mass models. In many models, the same fields that participate in the neutrino mass mechanism also provide viable dark matter candidates, thereby unifying two major shortcomings of the SM into the same framework.

Despite their attractiveness, minimal scotogenic realizations typically rely on small lepton number violating interactions where the smallness of the parameters, although allowed, is often introduced arbitrarily. For example, the scalar mass splitting that controls the neutrino mass scale in the one-loop neutrino mass models \cite{Balakrishna:1988ks, Ma:1988fp, Ma:1989ys, Ma:1990ce, Ma:1998dn, Tao:1996vb, Ma:2006km, Gu:2007ug, Ma:2008cu, Hirsch:2013ola, Aranda:2015xoa, Restrepo:2015ura, Longas:2015sxk, Fraser:2015zed, Fraser:2015mhb, Wang:2015saa, Arbelaez:2016mhg, vonderPahlen:2016cbw, Nomura:2016emz, Kownacki:2016hpm, Nomura:2017emk, Nomura:2017vzp, Bernal:2017xat, Wang:2017mcy, Bonilla:2018ynb, Calle:2018ovc, Avila:2019hhv, CarcamoHernandez:2018aon, Alvarado:2021fbw, Mandal:2021yph, Arbelaez:2022ejo, Cepedello:2022xgb, CarcamoHernandez:2022vjk, Leite:2023gzl,Kumar:2025cte,Kumar:2025zvv} is one such parameter. Such a feature suggests that the corresponding effective parameter may itself originate from naturally suppressed effects, like quantum corrections, opening the possibility that the neutrino mass mechanism should be understood as effectively emerging at a higher loop order.

From this perspective, two-loop mechanisms are especially well motivated \cite{Bonilla:2016diq,Baek:2017qos,Saad:2019vjo,Nomura:2019yft,Arbelaez:2019wyz,Saad:2020ihm,Xing:2020ezi,Chen:2020ptg,Nomura:2020dzw,Bonilla:2023aij,CarcamoHernandez:2026gjz}. Besides providing an additional suppression for neutrino masses, they offer a natural framework in which a small effective mass splitting can emerge dynamically. More generally, this allows the neutrino mass scale to be tied to the structure of the extended scalar and fermion sectors in a non-trivial way.

The hierarchy of the charged fermion masses and mixings is also not addressed by the SM. Within this context, modular flavour symmetries provide an economical and predictive alternative to conventional flavour models since finite modular groups are isomorphic to discrete groups. Modular flavour symmetries have been implemented in several extensions of the SM to explain the observed pattern of fermion masses and mixing angles \cite{Feruglio:2017spp,Ding:2023htn,Penedo:2018nmg,Novichkov:2018ovf,Novichkov:2018yse,Novichkov:2018nkm,King:2019vhv,Okada:2019xqk,Liu:2019khw,Kobayashi:2019rzp,Novichkov:2019sqv,deMedeirosVarzielas:2019cyj,Du:2020ylx,Abbas:2020qzc,deMedeirosVarzielas:2020kji,deMedeirosVarzielas:2021pug,Novichkov:2021evw,deMedeirosVarzielas:2022ihu,Ishiguro:2022pde,Okada:2022kee,Chen:2023mwt,Meloni:2023aru,Kobayashi:2023qzt,deMedeirosVarzielas:2023ujt,deMedeirosVarzielas:2023crv,Kumar:2023moh,Penedo:2024gtb,Ding:2024fsf,Belfkir:2024wdn,Ding:2024inn,Marciano:2024nwm,Nomura:2024ctl,Nomura:2024atp,Ding:2024ozt,Pathak:2024sei,Kobayashi:2025hnc,Kumar:2025nut,Kumar:2025lkv,Abhishek:2025ety,CarcamoHernandez:2025atu,Tavartkiladze:2025oiq,Nomura:2025raf,Chen:2025tby,Jangid:2025thp,Gao:2025jlw,Arriaga-Osante:2025ppz,Li:2025kcr,Behera:2025tpj,Nanda:2025lem,Zhang:2025dsa,Knapp-Perez:2025cht,Priya:2025wdm,Granelli:2025lds}. In these modular flavour models, the Yukawa couplings are promoted to modular forms and the resulting fermion mass matrix textures are dictated by modular group invariance. In this way, flavour texture and neutrino phenomenology can be directly linked to the spontaneous breaking of the modular symmetry, reducing arbitrariness of the fermion sector parameters.

In this work, we consider a modular $\Gamma_4\simeq S_4$ flavour symmetry supplemented by a discrete $Z_3$ symmetry. After the $\tau-$modulus acquires a vacuum expectation value (VEV) that triggers the spontaneous breaking of the modular $S_4$ symmetry, a preserved $Z_2$ parity symmetry remains after symmetry breaking. The inert scalars and right-handed neutrinos in the model carry odd modular weights, which corresponds to non-trivial $Z_2$ parities. This residual symmetry plays a central role: it forbids unwanted lower-order contributions, ensuring the radiative origin of neutrino masses, and stabilizes the $Z_2-$odd states that may account for the observed dark matter relic abundance. We employ $S_4$ because it is the smallest non-Abelian finite group containing singlet, doublet, and triplet irreducible representations. It is worth mentioning that the modular $S_4$ group has received a lot of attention by the particle physics community (see for instance Ref.~\cite{King:2019vhv,Kobayashi:2019mna,Wang:2020dbp,Zhao:2021jxg,Ding:2019gof,Novichkov:2020eep,Penedo:2018nmg,King:2021fhl,deMedeirosVarzielas:2019cyj,Wang:2019ovr,Novichkov:2018ovf} for several $\Gamma_4\simeq S_4$ flavour models).  

A distinctive feature of this model is that the discrete and modular charge assignments forbid the scalar interaction that would generate neutrino masses at one-loop level, namely the analogue of the usual scotogenic $\lambda_5(\phi^\dagger\eta)^2$ operator, while the inert CP-even and CP-odd scalar sectors remain degenerate at tree-level. The required mass splitting is instead radiatively generated, so that the active neutrino masses arise from an effective scotogenic mechanism at the two-loop level as a consequence of the symmetries of the model. 

 The same Yukawa interactions participate in neutrino mass generation and in charged lepton flavour violation processes, and the model predicts observable rates for the LFV processes that could be tested in future experiments \cite{MEGII:2025gzr,Belle:2021ysv,Hayasaka:2012pj}. In addition, the remnant $Z_2$ parity allows for scalar and fermionic dark matter candidates, both with qualitatively different predictions for relic abundance and direct detection observations depending on the singlet-doublet composition of the lightest dark state. In particular, the tree-level degeneracy of the CP-even and CP-odd inert states gives rise to an effectively complex scalar DM candidate before radiative splitting effects are included. This leads a rich phenomenology associated with its singlet-doublet mixture, and a fermionic candidate strongly correlated with LFV and neutrino masses.
In this work we study the viable scalar and fermionic dark matter regimes, identify the leading experimental constraints, and highlight the regions of parameter space that remain testable in future searches.

The paper is organized as follows. In Sec.~\ref{sec:model} we present the model symmetry and field content. In Sec.~\ref{sec:scalarsector} we discuss the scalar sector and the singlet-doublet mixture of the neutral inert scalar spectrum. Sec.~\ref{sec:lepton} is devoted to the neutrino mass mechanism and a discussion on the numerical analysis performed to fit the experimental data. Sec.~\ref{sec:lfv} is devoted to charged lepton flavour violation processes, while Sec.~\ref{sec:dark_matter} contains the dark matter analysis. We summarize our results in Sec.~\ref{sec:summary}.

%%%%%%%%%%%%%%%%%%%%

\section{The Model}
\label{sec:model}

Our model extends the Standard Model by an inert scalar doublet, two scalar singlets, and four right-handed neutral fermions. The symmetry group is enlarged by a finite modular flavour symmetry $\Gamma_4\simeq S_4$ and by an additional discrete $Z_3$ symmetry. The new fermionic sector consists of three right-handed neutrinos $N_{1R,2R,3R}$, arranged in an $S_4$ triplet, and an additional singlet fermion $\Omega_R$. The scalar sector consists of one $SU(2)_L$ scalar doublet $\eta$ and two gauge-singlet scalars $\sigma$ and $\varphi$.
 
The scalar fields are written as follows,
\begin{equation}
\phi =\left( 
\begin{array}{c}
G^{+} \\ 
\frac{1}{\sqrt{2}}\left(v+\phi_R+iG_0\right)
\end{array}
\right), \ \ \ 
\eta =\left( 
\begin{array}{c}
\eta ^{+} \\ 
\frac{1}{\sqrt{2}}(\eta_R+i\eta_I)
\end{array}
\right), \ \ \ 
\sigma =\frac{1}{\sqrt{2}}(v_{\sigma }+\sigma_{R}+iA), \ \ \ \varphi =%
\frac{1}{\sqrt{2}}(\varphi_{R}+i\varphi_{I}),
\label{eq: scalar_fields}
\end{equation}
where $\phi$ is the SM Higgs doublet. Here, $v\approx 246$ GeV.
\begin{table}[]
\centering
\begin{tabular}{c c c c c c c c c}
\toprule[0.2mm]
\hline
\vspace{0.1cm}   & $\phi$ & $\sigma$ & $\eta$ & $\varphi$ & $\overline{l}_{L}$ & $e_{R}$ & $N_{R}$ & $\Omega _{R}$ \\  
\midrule[0.1mm]
\hline
$SU(3)_C$ & $\mathbf{1}$ & $\mathbf{1}$ & $\mathbf{1}$ & $\mathbf{1}$ & $\mathbf{1}$ & $\mathbf{1}$ & $\mathbf{1}$ & $\mathbf{1}$ \\
$SU(2)_L$ & $\mathbf{2}$ & $\mathbf{1}$ & $\mathbf{2}$ & $\mathbf{1}$ & $\mathbf{2}$ & $\mathbf{1}$ & $\mathbf{1}$ & $\mathbf{1}$ \\
$U(1)_Y$ & $1/2$ & $0$ & $1/2$ & $0$ & $1/2$ & $-1$ & $0$ & $0$ \\
 $S_4$ & $\mathbf{1}$ & $\mathbf{1}$ & $\mathbf{1}$ & $\mathbf{1}$ & $\mathbf{3}$ & $\mathbf{3}$ & $\mathbf{3}$ & $\mathbf{1}^{\prime}$ \\
     $k$ & $0$ & $0$ & $-1$ & $1$ & $2$ & $2$ & $3$ & $2$ \\
    $Z_3$ & $0$ & $-1$ & $1$ & $0$ & $0$ & $0$ & $1$ & $-1$ \\
\bottomrule[0.2mm]
\end{tabular}
\caption{Particle content of the model and its assignments under the SM gauge group, modular $S_4$ and $Z_3$ discrete symmetry.}
\label{field_assignments}
\end{table}

In our model, the $\tau$-modulus and the $\sigma$ singlet acquire a VEV, driving a spontaneous breaking of both the $S_4$ and the $Z_3$ discrete symmetries. We assume that the breaking of the modular $S_4$ group yields a remnant $Z_2$ symmetry, under which scalar and fermionic particles inherit a charge $Q_{Z_2}=(-1)^k$, where $k$ is the modular weight of the particle under consideration. The heavy Majorana neutrinos also obtain their masses via Yukawa interactions with $\sigma$. The model particle content and the corresponding assignments under the $S_4\times Z_3$ discrete group are displayed in Table \ref{field_assignments}.

The preserved $Z_2$ symmetry serves a dual purpose: it guarantees the radiative origin of active neutrino masses, thereby setting the leading contribution at the two-loop level, and stabilizes the lightest state charged under $Z_2$, providing a dark matter candidate. The $Z_3$ symmetry, in turn, forbids the quartic scalar interaction $Y_{\mathbf{1}}^{\left(-2\right) }(\tau)\left(\phi^{\dagger}\eta\right)^2 $ that could yield a one-loop level contribution to active neutrino masses.

The symmetry breaking pattern of the full model is:
\begin{eqnarray}
&&\mathcal{G}=SU(3)_{C}\times SU\left( 2\right) _{L}\times U\left( 1\right)
_{Y}\times S_4\times Z_3 {\xrightarrow{v_\sigma,\braket{\tau}}}   \notag \\
&&\hspace{7mm}SU(3)_{C}\times SU\left( 2\right) _{L}\times U\left( 1\right)
_{Y}\times Z_2 {\xrightarrow{v}}   \notag \\
&&\hspace{7mm}SU(3)_{C}\times U\left( 1\right) _{Q}\times Z_2.  \label{SB}
\end{eqnarray}

\begin{figure}
    \centering
    \includegraphics[width=0.35\linewidth]{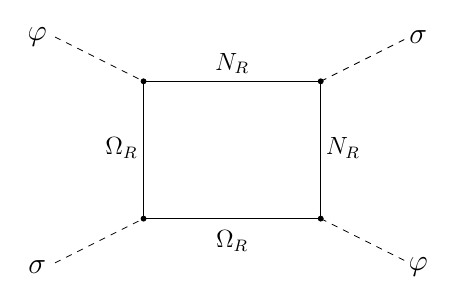}
    \caption{Scalar box diagram that induces the effective $\varphi^2$ coupling after the heavy neutrinos are integrated out and $\sigma$ acquires a VEV.}
    \label{fig:box}
\end{figure}
As a consequence of the non-trivial modular weight of the scalar fields, the $\tilde{\mu}^{ 2}_{\varphi} \varphi^2$ quadratic term is forbidden by the absence of a trivial $S_4$-singlet modular form of weight $2$. This implies the absence of tree-level mass splitting between CP-even and CP-odd physical scalar states, which has direct impact on the DM phenomenology. However, the discrete symmetry breaking allows this term to be generated radiatively through the box diagram of Fig.~\ref{fig:box}, in which the $\varphi^2\sigma\sigma^\ast$ effective interaction appears after the heavy neutrinos are integrated out.

\begin{figure}[]
    \centering
   \includegraphics[width=0.5\linewidth]{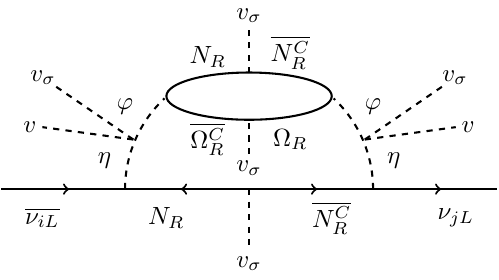}
    \caption{Two-loop radiative seesaw mechanism for the active neutrino sector.
    }
    \label{fig:Twoloopneutrinos}
\end{figure}

Fig.~\ref{fig:Twoloopneutrinos} displays the two-loop mechanism responsible for active neutrino mass generation. Here, the tree-level $\left(\phi^{\dagger}\eta\right)\varphi\sigma$ quartic interaction and the box interaction of Fig.~\ref{fig:box} discussed above make the two-loop diagram possible after the spontaneous breaking of the discrete symmetries and electroweak symmetry breaking.

The most general Yukawa interactions consistent with the symmetries of the model and the particle content shown in Table~\ref{field_assignments} are:
\begin{equation}
\label{eq:leptonYukawa}
-\mathcal{L}_{Y}^{\left( l\right) }=y_{1}^{\left( l\right) }Y_{\mathbf{1}%
}^{\left( 4\right) }\left( \tau \right) \left( \overline{l}_{L}\phi
e_{R}\right) _{\mathbf{1}}+y_{2}^{\left( l\right) }Y_{\mathbf{2}}^{\left(
4\right) }\left( \tau \right) \left( \overline{l}_{L}\phi e_{R}\right) _{%
\mathbf{2}}+y_{3}^{\left( l\right) }Y_{\mathbf{3}}^{\left( 4\right) }\left(
\tau \right) \left( \overline{l}_{L}\phi e_{R}\right) _{\mathbf{3}%
}+y_{4}^{\left( l\right) }Y_{\mathbf{3}^{\prime }}^{\left( 4\right) }\left(
\tau \right) \left( \overline{l}_{L}\phi e_{R}\right) _{\mathbf{3}^{\prime
}}+ \mathrm{H.c.} ,
\end{equation}
\begin{eqnarray}
-\mathcal{L}_{Y}^{\left( \nu \right) } &=&y_{1}^{\left( \nu \right) }Y_{%
\mathbf{1}}^{\left( 4\right) }\left( \tau \right) \overline{l}_{L}\eta
N_{R}+y_{2}^{\left( \nu \right) }Y_{\mathbf{2}}^{\left( 4\right) }\left(
\tau \right) \overline{l}_{L}\eta N_{R}+y_{3}^{\left( \nu \right) }Y_{%
\mathbf{3}}^{\left( 4\right) }\left( \tau \right) \overline{l}_{L}\eta
N_{R}+y_{4}^{\left( \nu \right) }Y_{\mathbf{3}^{\prime }}^{\left( 4\right)
}\left( \tau \right) \overline{l}_{L}\eta N_{R}  \notag \\
&&+y_{1}^{\left( N\right) }Y_{\mathbf{1}}^{\left( 6\right) }\left( \tau
\right) N_{R}\sigma ^{\ast }\overline{N_{R}^{C}}+y_{2}^{\left( N\right) }Y_{%
\mathbf{2}}^{\left( 6\right) }\left( \tau \right) N_{R}\sigma ^{\ast }%
\overline{N_{R}^{C}}+y_{3}^{\left( N\right) }Y_{\mathbf{3}}^{\left( 6\right)
}\left( \tau \right) N_{R}\sigma ^{\ast }\overline{N_{R}^{C}}+y_{4}^{\left(
N\right) }Y_{\mathbf{3}^{\prime }}^{\left( 6\right) }\left( \tau \right)
N_{R}\sigma ^{\ast }\overline{N_{R}^{C}}  \notag \\
&&+x_{\Omega }Y_{\mathbf{3}^{\prime }}^{\left( 6\right) } (\tau)
N_{R}\varphi \overline{\Omega _{R}^{C}}
+y_{\Omega }Y_{\mathbf{1}}^{\left(
4\right) } (\tau)\Omega _{R}\sigma \overline{\Omega _{R}^{C}}+\mathrm{H.c.},
\label{eq:nu-N-Om}
\end{eqnarray}
here $Y_{\mathbf{X}}^{(k)}(\tau)$ denote modular forms of weight $k$ in the irreducible representation $\mathbf{X}$ of $S_4$~\cite{Novichkov:2018ovf}, which are holomorphic functions of the modulus $\tau$. Explicit expressions for the modular forms as functions of the modulus can be found in Appendix \ref{app:modularforms}.

%%%%%%%%%%%%%%%%%%%%

\section{The scalar sector}
\label{sec:scalarsector}

According to the particle field assignments summarized in Table~\ref{field_assignments}, the scalar sector consists of two $SU(2)_L$ scalar doublets with identical hypercharges $Y=1/2$: the active Higgs doublet $\phi\sim(\mathbf{2},1/2)$ and the inert doublet $\eta\sim(\mathbf{2},1/2)$. In addition, the model contains two $SU(2)_L$ singlet scalars, namely the active complex field $\sigma\sim(\mathbf{1},0)$ and the inert scalar $\varphi\sim(\mathbf{1},0)$. From these fields, we construct the most general renormalizable scalar potential invariant under the $SU(2)_L\times U(1)_Y\times S_4 \times Z_3$ symmetry, which accommodates electroweak symmetry breaking, preserves the stability of the dark matter candidate, and is consistent with modular invariance. The resulting scalar potential reads
\begin{equation}
\label{eq:scalarpotential}
\begin{aligned}
V = & \ - \mu_\phi^2 \phi^\dagger \phi + \mu_\eta^2\eta^\dagger \eta + \mu_\sigma^2 \sigma^\dagger \sigma
+\mu^2_\varphi \varphi^\dagger\varphi 
 +\lambda_1 (\phi^\dagger \phi)^2 + \lambda_2 (\eta^\dagger \eta) ^2 + \lambda_3 (\sigma^\dagger \sigma)^2  
+\lambda_4 (\varphi^\dagger \varphi)^2
\\
& + \lambda_5(\phi^\dagger \eta) (\eta^\dagger \phi) + \lambda_6(\phi^\dagger \phi)(\eta^\dagger \eta) + \lambda_7 (\phi^\dagger \phi)(\varphi^\dagger \varphi)+ \lambda_8 (\phi^\dagger \phi)(\sigma^\dagger \sigma) + \lambda_9 (\eta^\dagger \eta)(\varphi^\dagger \varphi) \\
& + \lambda_{10} (\eta^\dagger \eta)(\sigma^\dagger \sigma)
+\lambda_{11}(\varphi^\dagger \varphi)(\sigma^\dagger \sigma)
+\left[ \lambda_{12}   (\phi^\dagger \eta) \sigma \varphi + \lambda_{13}Y_{\mathbf{1}}^{(4)}(\tau) \varphi^4  + \kappa \sigma^3+\mathrm{H.c.} \right] ,
\end{aligned}
\end{equation}
where $\mu_i$ denote the scalar mass parameters, $\lambda_i$ are dimensionless quartic couplings, and $\kappa$ is a trilinear scalar coupling.

In the scalar potential, quadratic operators are not written with explicit modular-dependent factors. For a scalar field $\Phi$ with modular weight $k$, the modular-invariant kinetic and quadratic terms contain the factor $(-i\tau+i\bar{\tau})^{-k}$, which becomes a constant once the modulus $\tau$ is fixed.
This factor can be absorbed into the canonical normalization of $\Phi$ and into the corresponding effective mass parameter. Consequently, all allowed bilinear terms $\Phi^\dagger\Phi$ are written without explicit modular factors, while the residual $\tau$ dependence is encoded in the
effective scalar couplings. This applies to $\Phi=\phi,\ \sigma,\ \eta,$ and $\varphi$.

The scalar potential minimization conditions are
\begin{equation}
\begin{aligned}
\mu_\phi ^2 =& \ \lambda_1 v^2 + \frac{1}{2} \lambda_8 v_{\sigma}^2,\\
\mu_\sigma ^2 =& \ -\frac{3}{\sqrt{2}} \kappa v_\sigma - \lambda_3 v_\sigma^2 - \frac{1}{2} \lambda_8 v^2.
\end{aligned}
\end{equation}

After SSB, the Goldstone bosons associated with the breaking of $SU(2)_L \times U(1)_Y$ are absorbed as the longitudinal components of $W^\pm$ and $Z$ gauge bosons. In the CP-even neutral visible sector, the remaining physical scalar degrees of freedom originate from the mixing between the active Higgs doublet and the scalar singlet.

In the CP-even neutral sector, the fields $\phi_R$ and $\sigma_R$ mix and the physical states are obtained through the rotation
\begin{equation}
\binom{h}{H}=
\begin{pmatrix}
\cos\alpha & \sin\alpha\\
-\sin\alpha & \cos\alpha
\end{pmatrix}
\binom{\phi_R}{\sigma_R},
\end{equation}
where $h$ is identified with the SM-like Higgs boson, while $H$ corresponds to an additional CP-even scalar state and the mixing angle satisfies
\begin{equation}
\tan( 2\alpha)  =   -\frac{4 v v_\sigma \lambda_{8}}
       {3\sqrt{2} \kappa v_\sigma - 4 v^{2} \lambda_{1} + 4 v_\sigma^{2} \lambda_{3}}.
\label{eq:alpha}
\end{equation}

In the CP-odd sector, there are two states coming from the imaginary part of the neutral component of $\phi$ and $\sigma$ in Eq.~\eqref{eq: scalar_fields}, they correspond to the SM Nambu-Goldstone boson $G_0$ associated with the longitudinal polarization of the $Z$ gauge boson and to a physical massive pseudoscalar $A$ with squared mass
\begin{equation}
    m^2_A= - \frac{9}{\sqrt{2}} \kappa v_\sigma, 
\end{equation}
where the requirement for positive squared mass imposes $\kappa<0$. In addition, the electrically charged component of the $\phi$ doublet, namely $G^\pm$ corresponds to the SM Nambu-Goldstone boson associated with the longitudinal polarization of $W^\pm$ gauge boson.

Moving to the inert sector, the quadratic interaction $\varphi^2$ is forbidden by the modular $S_4$ symmetry at tree-level due to the absence of a trivial $S_4$ singlet modular form of weight 2. This term would provides opposite contributions to the CP-even and CP-odd squared mass matrices. Since this term is forbidden, masses of the neutral inert scalars are identical at tree-level. The interaction is radiatively generated through the inner loop in Fig.~\ref{fig:Twoloopneutrinos}. The induced splitting is small and will be included only in the neutrino mass and DM phenomenology discussed in Secs.~\ref{sec:lepton} and \ref{sec:lepton} respectively.

The neutral scalar spectrum of the inert sector can be assumed  consisting of two neutral massive complex scalars, $\chi_{1,2}=(S_{1,2}+iP_{1,2})/\sqrt{2}$, where $S_{1,2}$ and $P_{1,2}$ denote the CP-even and CP-odd states in the mass-diagonal basis. The inert physical states are given by

\begin{equation}
\label{eq:physinert}
\begin{aligned}
\begin{pmatrix}
    \chi_2\\
    \chi_1
\end{pmatrix}=\begin{pmatrix}
    \cos{\beta} & \sin{\beta} \\
    -\sin{\beta} & \cos{\beta}
\end{pmatrix}\begin{pmatrix}
    \eta_R\\
    \varphi_R
\end{pmatrix},
\end{aligned}
\end{equation}
with
\begin{equation}
\begin{aligned}
    \tan( 2\beta)=\frac{2vv_\sigma
    \lambda_{12}}{v_\sigma^2(\lambda_{10}-\lambda_{11})+v^2(\lambda_5+\lambda_6-\lambda_7)+2
    (
    \mu_\eta^2
    -\mu_\varphi^2)
    },
\label{eq:beta}
\end{aligned}
\end{equation}
the CP-even mixing angle\footnote{Although the CP-even and CP-odd mass matrices have identical eigenvalues at tree level, the mixing angles that relate the interaction and mass eigenstates have opposite signs, $\beta_{\rm odd}=-\beta$.}. 
Note that we are using the convention from Ref.~\cite{Beniwal:2020hjc}, where in the limiting case $\beta=0$, the lightest inert scalar is identified with the singlet state. In addition, there is an electrically charged scalar state $\eta^\pm$ originating from the $\eta$ doublet in Eq.~\eqref{eq: scalar_fields}.
The squared masses of the inert scalars are
\begin{equation}\label{eq:mass}
\begin{aligned}
m_{\chi_{1,2}}^{2}
=   & \ 
\frac{1}{2}
\left[
A_{\eta\eta}+A_{\varphi\varphi}
\mp
\sqrt{
\left(A_{\eta\eta}-A_{\varphi\varphi}\right)^{2}
+4A_{\eta\varphi}^{2}}
\right] ,\\
m^2_{\eta^\pm}= & \ \mu_\eta^2+\frac{\lambda_6 v^2}{2}+\frac{\lambda_{10} v_\sigma^2}{2},
\end{aligned}
\end{equation}
with
\begin{equation}
\begin{aligned}
A_{\eta\eta}&=
\mu_\eta^2+\frac{1}{2}\Big[(\lambda_5+\lambda_6)v^2+\lambda_{10}v_\sigma^2\Big],\\
A_{\varphi\varphi} &=
\mu_\varphi^2+\frac{1}{2}\Big[\lambda_7 v^2+\lambda_{11}v_\sigma^2\Big],\\
A_{\eta\varphi} &= \frac{1}{2}\lambda_{12}vv_\sigma.
\end{aligned}
\end{equation}

\subsection{Constraints on the scalar potential}

The parameters of the scalar potential in Eq.~\eqref{eq:scalarpotential} are subject to theoretical constraints. 
In particular, perturbative unitarity requires the quartic couplings to remain within the perturbative regime, which is typically implemented through the condition $|\lambda_i|<4\pi$. 
Furthermore, vacuum stability demands that the scalar potential be bounded from below. 
Since this property is governed by the highest-dimensional operators in the large-field regime, we focus on the quartic part of the potential, neglecting quadratic and trilinear terms. 
In the large-field limit, the relevant contribution can be written as
\begin{equation}
\begin{aligned}
V_4 \simeq & \ \left(  \sqrt{\lambda_1} a - \sqrt{\lambda_2} b\right)^2  + \left(  \sqrt{\lambda_3} c - \sqrt{\tilde{\lambda}_4} d\right)^2 
+ \left(2\sqrt{\lambda_1 \lambda_2}+\lambda_6 \right)\left[ ab-f^2-e^2\right]\\
& +\left(2\sqrt{\lambda_1 \lambda_2} +\lambda_5 + \lambda_6 \right) \left[ f^2+e ^2\right]+
\left(2\sqrt{\lambda_3 + \tilde{\lambda}_4} +\lambda_{11}\right)cd +
\lambda_{7}ad+ \lambda_8 ac  + \lambda_9 bd + \lambda_{10} bc,
\end{aligned}
\end{equation}
where we have introduced the gauge-invariant quantities 
$a \equiv \phi^\dagger \phi$, 
$b \equiv \eta^\dagger \eta$, 
$c \equiv \sigma^\dagger \sigma$, 
$d \equiv \varphi^\dagger \varphi$, 
$e \equiv \mathrm{Im}(\phi^\dagger \eta)$, and 
$f \equiv \mathrm{Re}(\phi^\dagger \eta)$, 
with $\tilde{\lambda}_4 \equiv \lambda_4 - 2 \left| \lambda_{13} Y_{\mathbf{1}}^{(4)}(\tau) \right|$.

Following the standard procedure for the stability analysis~\cite{Gunion:2002zf,Bhattacharyya:2015nca,Maniatis:2006fs}, we obtain a sufficient set of bounded-from-below conditions for the scalar potential:
\begin{equation}
\begin{aligned}
& \lambda_{1,2,3},\,\tilde{\lambda}_4 >0, \quad -2\sqrt{\lambda_1 \lambda_2} <\lambda_6 , \lambda_5 + \lambda_6 , \quad -2 \sqrt{\lambda_3 \tilde{\lambda}_4} <\lambda_{11},
\quad 2\sqrt{\lambda_1 \tilde{\lambda}_4} + \lambda_7 >0,\\
& 2\sqrt{\lambda_1 \lambda_3} + \lambda_8 >0, \quad 2\sqrt{\lambda_2 \tilde{\lambda}_4} + \lambda_9>0, \quad 2\sqrt{\lambda_2 \lambda_3} + \lambda_{10}>0.
\end{aligned}
\end{equation}

Additionally, a stringent constraint arises from the diphoton signal strength of the SM-like Higgs boson, measured with high precision by CMS~\cite{Saha:2022cnz} and ATLAS~\cite{ATLAS:2022tnm}. 
In the SM, the decay width of the Higgs boson into two photons is dominated by the interference between the $W$-boson and top-quark loops, while in our model the charged scalar $\eta^\pm$ induces an extra one-loop contribution controlled mainly by the quartic couplings $\lambda_{6,10}$. 

The numerical prediction of our model for the diphoton signal strength $R_{\gamma\gamma}$ imposes both an upper and a lower bound on the mixing angle in the visible CP-even sector, $0.990 \lesssim \cos(\alpha) \lesssim 1$. 
In our scenario, the Higgs coupling to electroweak gauge bosons is universally rescaled by the doublet fraction of the SM-like state, so that the bound derived from the diphoton signal strength directly translates into a stringent constraint on the vector-boson coupling modifier and, consequently, on the mixing angle of the visible CP-even sector, in agreement with global LHC coupling fits requiring $\kappa_V = 1.035 \pm 0.031$~\cite{ATLAS:2022vkf}.

\begin{figure}[h]
\centering
\subfigure[]{\includegraphics[width=0.49\linewidth]{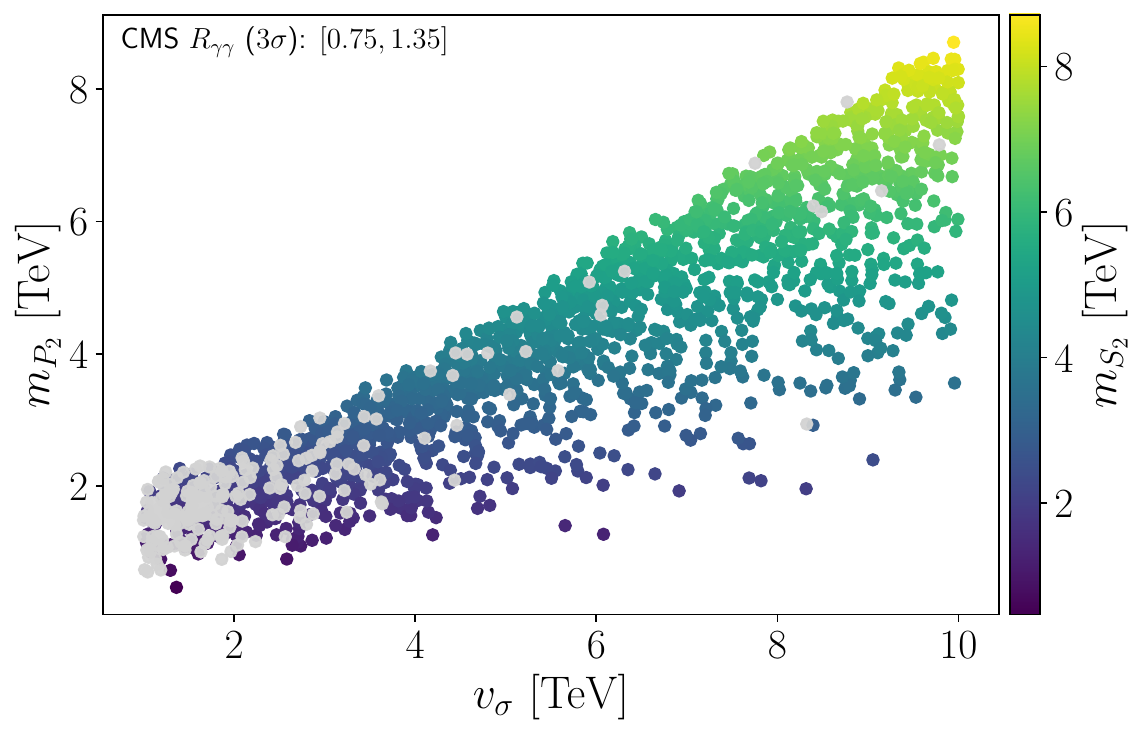}}
\subfigure[]{\includegraphics[width=0.49\linewidth]{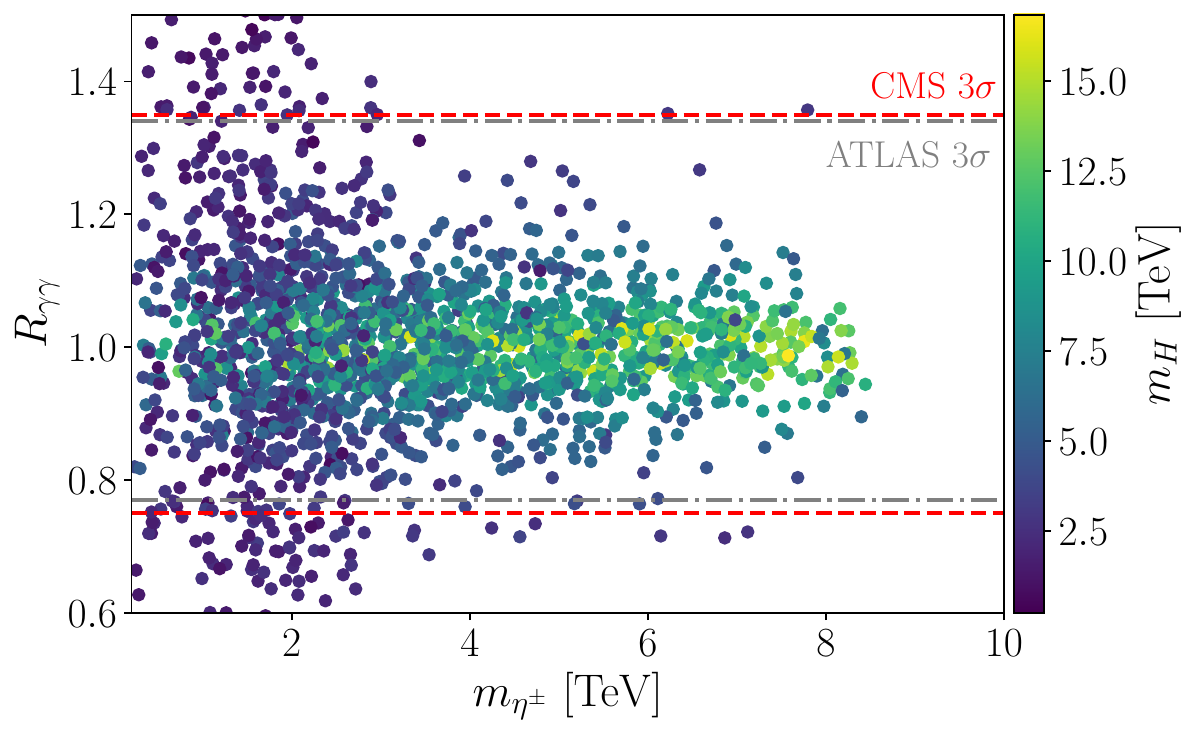}}
\caption{
(a) Allowed parameter space compatible with the CMS $3\sigma$ interval for $R_{\gamma\gamma}$ in the plane $m_{P_2}$-$v_\sigma$, with the color bar indicating the scalar mass $m_{S_2}$ in TeV. 
(b) Dependence of $R_{\gamma\gamma}$ on the charged scalar mass $m_{\eta^\pm}$, where the color bar denotes the heavy visible scalar mass $m_H$ in TeV. 
Red dashed lines correspond to the CMS $3\sigma$ range, while gray dot-dashed lines indicate the ATLAS $3\sigma$ interval.}

\label{fig:Rgammagamma}
\end{figure}

In Fig.~\ref{fig:Rgammagamma}, panel (a), we display the region of parameter space compatible with the diphoton signal strength $R_{\gamma\gamma}$ within the $3\sigma$ range reported by CMS. The horizontal axis corresponds to the VEV of the active singlet, $v_\sigma$, while the vertical axis represents the mass of the heaviest inert pseudoscalar, $m_{P_2}$, and the color bar indicates the mass of the scalar field $m_{S_2}$. An approximately linear correlation between $m_{P_2}$ and $v_\sigma$ is observed, which can be understood from the fact that the inert scalar masses receive dominant contributions proportional to $\lambda_i\, v_\sigma^2$. Consequently, larger values of $v_\sigma$ shift the scalar spectrum toward heavier masses. Regions with smaller values of $v_\sigma$ are strongly constrained by the SM-like Higgs signal. All colored points satisfy the experimental constraint on $R_{\gamma\gamma}$ within the CMS $3\sigma$ allowed interval, whereas the gray points correspond to configurations excluded by this bound.

On the other hand, Fig.~\ref{fig:Rgammagamma}, panel (b) shows the dependence of $R_{\gamma\gamma}$ on the charged scalar mass $m_{\eta^\pm}$, where the color bar indicates the mass of the heavy visible scalar $H$ in TeV. The red dashed lines delimit the $3\sigma$ range allowed by CMS, while the gray dot-dashed lines correspond to the interval reported by ATLAS. It is evident that most of the viable points cluster around $R_{\gamma\gamma} \simeq 1$, with small deviations induced by the one-loop contributions of the charged scalar. For larger charged scalar masses, the additional contribution progressively decouples, and the prediction approaches the Standard Model value.

\subsection{Physical parametrization in terms of scalar mass splittings}

The composition of the scalars $\chi_{1,2}$ as mixtures of a doublet and a singlet is a key ingredient in the phenomenology of the scalar sector and of the scalar dark matter candidate. The different mixtures are determined by the value of the mixing angle $\beta$ in Eq.~(\ref{eq:beta}). For the lightest scalar $\chi_1$, the possible mixtures range from a mostly singlet-like state for $\beta\sim 0$ to a mostly doublet-like state for $\beta\sim \pi/2$ and conversely the heavier scalar $\chi_2$.

Given its phenomenological relevance, in this section we parametrize the inert scalar sector in terms of the mass splittings among the inert scalars, since, as we will show below, the ratio between these splittings determines the value of the mixing angle. We now replace the scalar-potential parameters by physical squared mass differences.
We begin by defining the squared mass splittings
\begin{equation}
\begin{aligned}
\Delta_0^2\equiv m_{\chi_2}^2-m_{\chi_1}^2,\qquad \Delta_\pm^2\equiv m_{\eta^\pm}^2-m_{\chi_1}^2.
\end{aligned}
\end{equation}
where $\chi_1$ is the lightest scalar.
From Eq.~(\ref{eq:mass}), we find  $\mu_\eta^2$ and $\mu_\varphi^2$ in terms of $m_{\chi_1}$ and $m_{\eta^\pm}$. This allows us to re-express the coupling $\lambda_{12}$ in terms of the mass splittings. Assuming $\Delta_0^2,\Delta_\pm^2>0$ and $\lambda_{12}\in\mathbb{R}^+$, the possible solutions are
\begin{equation}
\begin{aligned}
\lambda_{12}=
\begin{cases}
    \dfrac{1}{vv_\sigma}\sqrt{\Delta_t}\sqrt{2\Delta_0^2-\Delta_t} & \text{if }\Delta_t>0,\\
    \\
    \dfrac{1}{vv_\sigma}\sqrt{-\Delta_t}\sqrt{2\Delta_0^2+\Delta_t} & \text{if }\Delta_t<0,
\end{cases}
\end{aligned}
\end{equation}
where we have defined the mixture-scale parameter $\Delta_t=2\Delta_\pm^2+v^2\lambda_5$. The reality of $\lambda_{12}$ restricts the physical domain to $0\leq \Delta_t\leq 2\Delta_0^2$ in the branch $\Delta_t>0$, and to $-2\Delta_0^2\leq \Delta_t\leq 0$ in the branch $\Delta_t<0$.

After substituting these expressions into Eq.~(\ref{eq:beta}) and rearranging the result, we obtain
\begin{equation}
\begin{aligned}
\label{eq:sinbeta}
\sin{\beta}=
\begin{cases}
    \sqrt{1-\dfrac{\Delta_t}{2\Delta_0^2}} & \text{if }\Delta_t>0,\\
    \\
    \sqrt{-\dfrac{\Delta_t}{2\Delta_0^2}} & \text{if }\Delta_t\leq 0.
\end{cases}
\end{aligned}
\end{equation}

Expressing the mixing angle in this form confirms the usefulness of introducing $\Delta_t$. This parameter defines a characteristic scale for the mixing of the inert scalar sector. The ratio between $\Delta_t$ and $\Delta_0^2$ fixes the angle $\beta$, thereby determining whether $\chi_1$ is predominantly doublet-like or singlet-like. On the other hand, the solution branch with $\lambda_{12}<0$ does not lead to new phenomenology, since it only reverses the convention used to define $\beta$ while leaving the phenomenology unchanged.

\begin{figure}
    \centering
    \includegraphics[width=0.5\linewidth]{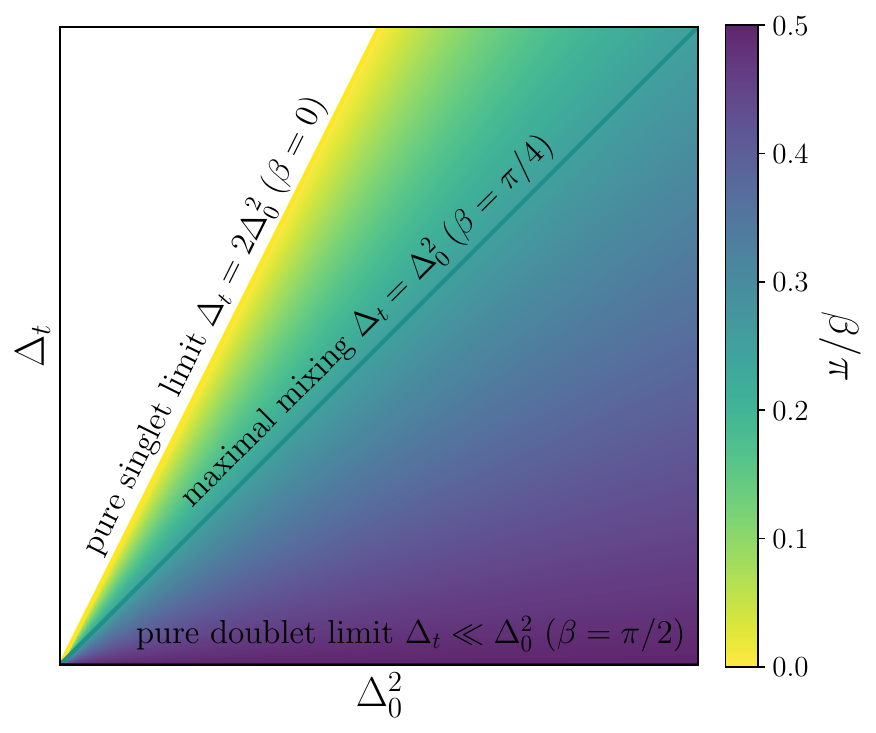}
    \caption{Color map for the mixture of the lightest inert neutral scalar in the $(\Delta_0^2,\Delta_t)$ plane, where $\Delta_0^2$ denotes the splitting between the two neutral scalars and $\Delta_t$ is the mixture-scale parameter. The color indicates the value of the mixing angle $\beta/\pi$. The lower region corresponds to $\chi_1$ being predominantly doublet-like, while the limit $\Delta_t\ll\Delta_0^2$ approaches the pure doublet case. The upper region corresponds to $\chi_1$ being predominantly singlet-like, and the boundary $\Delta_t=2\Delta_0^2$ defines the pure singlet limit. The intermediate line $\Delta_t=\Delta_0^2$ corresponds to maximal mixing, $\beta=\pi/4$. This diagram provides a geometric representation of the neutral-scalar mixtures in terms of physical mass splittings.}
    \label{fig:scalar_mixture}
\end{figure}

Note that the region with $\Delta_t\leq 0$ can only be reached if $\lambda_5<0$. This region defines the same mixing map as the one with $\Delta_t>0$. Without loss of generality, we can therefore focus on the region with $\lambda_5>0$. The $(\Delta_0^2,\Delta_t)$ plane shown in Fig.~\ref{fig:scalar_mixture} provides the mixture map of the inert scalar sector relevant to the phenomenology of $\chi_1$. In this plane, the limit $\Delta_t\ll \Delta_0^2$ corresponds to the case in which $\chi_1$ is purely doublet-like, whereas the boundary $\Delta_t=2\Delta_0^2$ defines the purely singlet-like case. Between these two extremes, the line $\Delta_t=\Delta_0^2$ defines the region where $\chi_1$ is a maximal doublet--singlet mixing. This parametrization will serve as the basis for the phenomenological analysis of the scalar dark matter scenario in Sec.~\ref{sec:dark_matter}.

%%%%%%%%%%%%%%%%%%%%

\section{Lepton masses and mixings}
\label{sec:lepton}

In this model, the charged-lepton and active neutrino mass matrices arise from the Yukawa interactions in Eqs.~\eqref{eq:leptonYukawa} and \eqref{eq:nu-N-Om}, respectively, after the spontaneous breaking of the electroweak and modular $S_4$ symmetries. In the charged-lepton sector, the vacuum expectation value of the modulus $\tau$ fixes the values of the modular forms, thereby determining the mass texture. The charged lepton mass matrix is given by:
\begin{equation}
M_{\ell}= \frac{v}{\sqrt{2}} \text{$\left( 
\begin{array}{ccc}
\text{{\scriptsize $y_{1}^{\left( l\right) }$\textsc{Y}$_{%
\mathbf{1}}^{\left( 4\right) }\left( \tau \right) $\textsc{+2$y$}$_{3}^{\left(
l\right) }$\textsc{Y}$_{\mathbf{3},1}^{\left( 4\right) }\left( \tau \right) $%
}} & \text{{\scriptsize \textsc{$y$}$_{2}^{\left( l\right) }$\textsc{Y}$_{%
\mathbf{2},1}^{\left( 4\right) }\left( \tau \right) $\textsc{-$y$}$%
_{3}^{\left( l\right) }$\textsc{Y}$_{\mathbf{3},3}^{\left( 4\right) }\left(
\tau \right) $\textsc{+$y$}$_{4}^{\left( l\right) }$\textsc{Y}$_{\mathbf{3}%
^{\prime },3}^{\left( 4\right) }\left( \tau \right) $}} & \text{{\scriptsize 
\textsc{$y$}$_{2}^{\left( l\right) }$\textsc{Y}$_{\mathbf{2},2}^{\left(
4\right) }\left( \tau \right) $\textsc{-$y$}$_{3}^{\left( l\right) }$\textsc{Y}%
$_{\mathbf{3},2}^{\left( 4\right) }\left( \tau \right) $\textsc{-$y$}$%
_{4}^{\left( l\right) }$\textsc{Y}$_{\mathbf{3}^{\prime },2}^{\left(
4\right) }\left( \tau \right) $}} \\ 
\text{{\scriptsize \textsc{$y$}$_{2}^{\left( l\right) }$\textsc{Y}$_{\mathbf{2}%
,1}^{\left( 4\right) }\left( \tau \right) $\textsc{-$y$}$_{3}^{\left( l\right)
}$\textsc{Y}$_{\mathbf{3},3}^{\left( 4\right) }\left( \tau \right) $\textsc{%
-$y$}$_{4}^{\left( l\right) }$\textsc{Y}$_{\mathbf{3}^{\prime },3}^{\left(
4\right) }\left( \tau \right) $}} & \text{{\scriptsize \textsc{$y$}$%
_{2}^{\left( l\right) }$\textsc{Y}$_{\mathbf{2},2}^{\left( 4\right) }\left(
\tau \right) $\textsc{+2$y$}$_{3}^{\left( l\right) }$\textsc{Y}$_{\mathbf{3}%
,2}^{\left( 4\right) }\left( \tau \right) $}} & \text{{\scriptsize \textsc{$y$}%
$_{1}^{\left( l\right) }$\textsc{Y}$_{\mathbf{1}}^{\left( 4\right) }\left(
\tau \right) $\textsc{-$y$}$_{3}^{\left( l\right) }$\textsc{Y}$_{\mathbf{3}%
,1}^{\left( 4\right) }\left( \tau \right) $\textsc{+$y$}$_{4}^{\left( l\right)
}$\textsc{Y}$_{\mathbf{3}^{\prime },1}^{\left( 4\right) }\left( \tau \right) 
$}} \\ 
\text{{\scriptsize \textsc{$y$}$_{2}^{\left( l\right) }$\textsc{Y}$_{\mathbf{2}%
,2}^{\left( 4\right) }\left( \tau \right) $\textsc{-$y$}$_{3}^{\left( l\right)
}$\textsc{Y}$_{\mathbf{3},2}^{\left( 4\right) }\left( \tau \right) $\textsc{+%
}$\text{{}}$\textsc{$y$}$_{4}^{\left( l\right) }$\textsc{Y}$_{\mathbf{3}%
^{\prime },2}^{\left( 4\right) }\left( \tau \right) $}} & \text{{\scriptsize 
\textsc{$y$}$_{1}^{\left( l\right) }$\textsc{Y}$_{\mathbf{1}}^{\left( 4\right)
}\left( \tau \right) $\textsc{-$y$}$_{3}^{\left( l\right) }$\textsc{Y}$_{%
\mathbf{3},1}^{\left( 4\right) }\left( \tau \right) $\textsc{-$y$}$%
_{4}^{\left( l\right) }$\textsc{Y}$_{\mathbf{3}^{\prime },1}^{\left(
4\right) }\left( \tau \right) $}} & \text{{\scriptsize \textsc{$y$}$%
_{2}^{\left( l\right) }$\textsc{Y}$_{\mathbf{2},1}^{\left( 4\right) }\left(
\tau \right) $\textsc{+2$y$}$_{3}^{\left( l\right) }$\textsc{Y}$_{\mathbf{3}%
,3}^{\left( 4\right) }\left( \tau \right) $}}%
\end{array}%
\right) $},
\end{equation}
where $\tau$ denotes the fixed vacuum value of the modulus. The charged-lepton mass matrix is diagonalized as
\begin{equation}
\begin{aligned}
U_\ell^\dagger M_\ell M_\ell^\dagger U_\ell
=
\mathrm{diag}(m_e^2,m_\mu^2,m_\tau^2).
\end{aligned}
\end{equation}

As can be seen from the field assignment in Table~\ref{field_assignments}, the modular $S_4$ symmetry forbids tree-level neutrino mass terms. However, the spontaneous breaking of the $S_4$ and $Z_3$ symmetries driven by $\braket{\tau}$ and $v_\sigma$, respectively, dynamically generates the Weinberg operator $l_Ll_L \phi\phi$ at two-loop level, then providing, after the electroweak symmetry breaking, the two-loop Majorana neutrino mass shown in Fig.~\ref{fig:Twoloopneutrinos}.

As we already commented in Sec.~\ref{sec:model},  
in our model the usual scotogenic mass splitting $(\sigma^2\varphi^2+\mathrm{H.c.})$ \cite{Ma:2006km} is forbidden at tree-level, but is generated at one-loop-level according to Fig.~\ref{fig:box} which is a subdiagram of the neutrino mass diagram in Fig.~\ref{fig:Twoloopneutrinos}. 
Although the underlying mechanism is truly two-loop, the resulting active neutrino mass matrix can be expressed in an effective scotogenic-like form:
\begin{equation}
\begin{aligned}
    \left(M_\nu\right)_{ij}\approx \frac{\lambda_\text{eff}v_\sigma^2}{32\pi^2}\sin^2\beta\cos^2\beta\sum_{k=1}^3\frac{\left(\tilde Y_{\nu }\right)_{ik} \left(\tilde Y_{\nu }\right)_{jk}}{m_{N_k}}\left[G(m_{\chi_1},m_{N_k})+G(m_{\chi_2},m_{N_k})\right],
\end{aligned}
\label{eq:actneutrino}
\end{equation}
where
\begin{equation}
    G(m,M)=\frac{M^2}{m^2-M^2}+\frac{M^4}{(m^2-M^2)^2}\ln\left(\frac{M^2}{m^2}\right),
\end{equation}
corresponds to the standard one-loop function \cite{Ma:2006km,Escribano:2020iqq}.
Meanwhile, $\widetilde{Y}_{\nu}=Y_{\nu}U_{N}$
is the matrix containing the couplings between active neutrinos and the physical heavy fermions $N_R$ whereas $U_N$ is the rotation matrix that diagonalizes the right-handed Majorana neutrino mass matrix $U_N^T M_N U_N=\mathrm{diag}(m_{N_1},m_{N_2},m_{N_3})$.
The effective coupling $\lambda_{\text{eff}}$ is induced at one loop through the box diagram of Fig.~\ref{fig:box}, as in Ref.~\cite{Chen:2020ptg}. This coupling has the form
\begin{equation}
\label{eq:lambdaeff}
    \lambda_\text{eff}=-\frac{1}{16\pi^2}\sum_{i,j=1}^3\left(\widetilde{Y}_N\right)_{ij}(\tilde x_\Omega)_j(y_\Omega)^2 J(m_{N_j}^2,m_\Omega^2),
\end{equation}
where $J(a,b)$ is the loop function that depends on the masses of the heavy fermions
\begin{equation}
    J(a,b)=\frac{ab}{(a-b)^3}\left[-2(a-b)+(a+b)\ln\left(\frac{a}{b}\right)\right],
\end{equation}
and $\widetilde{Y}_{N} = Y_N U_N$ is the matrix containing the couplings between havy fermions $N_R$ and $\sigma$. A complete derivation of Eqs.~\eqref{eq:actneutrino} and \eqref{eq:lambdaeff} as well as the full form for interaction matrices $\widetilde{Y}_{\nu}$ and $\widetilde{Y}_N$ can be found in Appendix \ref{app:neutrinomass}.

It is worth noting that the loop suppression is controlled by the inert scalars and the heavy neutral fermions. In particular, the same inert sector structure that determines the singlet--doublet composition of the neutral inert scalars also controls the size of the active-neutrino masses. 

The active neutrino mass matrix is diagonalized by
\begin{equation}
\begin{aligned}
U_\nu^T M_\nu U_\nu
=
\mathrm{diag}(m_{\nu_1},m_{\nu_2},m_{\nu_3}),
\end{aligned}
\end{equation}
where the mass-squared differences are defined as usual,
\begin{equation}
\begin{aligned}
\Delta m_{21}^2 = m_{\nu_2}^2 - m_{\nu_1}^2,
\qquad
\Delta m_{31}^2 = m_{\nu_3}^2 - m_{\nu_1}^2.
\end{aligned}
\label{eq:dmsq_new}
\end{equation}

The lepton mixing matrix is given by
\begin{equation}
\begin{aligned}
U_{\mathrm{PMNS}} = U_\ell^\dagger U_\nu,
\end{aligned}
\label{eq:PMNS}
\end{equation}
and the mixing angles are extracted from the mixing matrix as
\begin{equation}
\begin{aligned}
\sin^2\theta_{13} = |(U_{\mathrm{PMNS}})_{13}|^2,\hspace{.3cm}\sin^2\theta_{12} = \frac{|(U_{\mathrm{PMNS}})_{12}|^2}{1-| (U_{\mathrm{PMNS}})_{13}|^2},\hspace{.3cm}\sin^2\theta_{23} = \frac{|(U_{\mathrm{PMNS}})_{23}|^2}{1-| (U_{\mathrm{PMNS}})_{13}|^2}.
\end{aligned}
\label{eq:angles_new}
\end{equation}

Numerical results for the neutrino sector are analyzed by scanning the model parameters:
\begin{equation}
\begin{aligned}
\Bigl\{
\tau,y^{(l)},y^{(\nu)},y^{(N)},x_{\Omega},\lambda,m_{H},m_A,
m_{\chi_1},
m_{\Omega_R},v_\sigma,\Delta_\pm,\Delta_0\Bigr\},
\end{aligned}
\label{eq:scanbasis_new}
\end{equation}
where all leptonic Yukawa couplings are taken to be real. Assumptions over scalar quartic couplings will be discussed in Sec.~\ref{sec:dark_matter}.

For a given input vector, we minimize the standard $\chi^2$ function \cite{ParticleDataGroup:2024cfk} defined in terms of the calculated observables and their corresponding best experimental values arising from neutrino oscillation experimental data \cite{deSalas:2020pgw}. In addition, all accepted points are required to satisfy the cosmological bounds \cite{Jiang:2024viw,Naredo-Tuero:2024sgf}
\begin{equation}
\begin{aligned}
\sum_i m_{\nu_i} \leq 0.04-0.3~\mathrm{eV},
\end{aligned}
\end{equation}
the Higgs into di-photon rate $R_{\gamma\gamma}$, the SM vector boson coupling $\kappa_V$ \cite{ATLAS:2022tnm} and vacuum stability conditions.

In the numerical results presented below we take the input parameters in the following ranges:
\begin{equation}
\begin{aligned}
    &\hspace{0.5cm}\mathrm{Re}(\tau)\in [-1/2,1/2],\quad \mathrm{Im}(\tau)\in[0,4],\quad |\tau|\geq 1,\quad \left|y^{(l,\nu,N)}\right|,\left|x_\Omega\right|\in \left[0,\sqrt{4\pi}\right],\quad \left|\lambda\right|\in \left[0,4\pi\right],\\
    &m_H,m_A,m_{\chi_1}\in\left[10^2, 10^4\right]~\mathrm{GeV},\quad m_{\Omega_R}\in \left[2\times10^3,2\times10^4\right]~\mathrm{GeV},\quad v_\sigma\in\left[9\times10^3, 2.5\times10^4\right]~\mathrm{GeV}\\
    &\hspace{3cm}\Delta_\pm\in\left[20, 5\times10^3\right]~\mathrm{GeV},\quad \Delta_0\in\left[0.4\Delta_\pm, 1.5\Delta_\pm\right]~\mathrm{GeV}.
\end{aligned}
\end{equation}

\begin{figure}
\begin{center}
\includegraphics[width=0.485\textwidth]{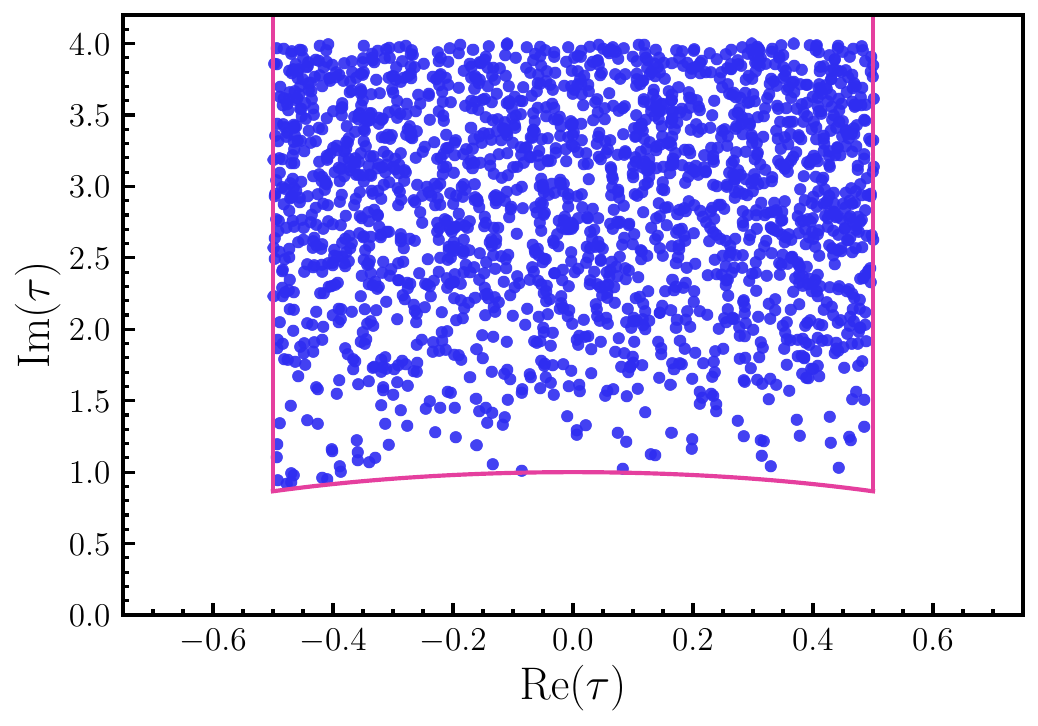}~~
\includegraphics[width=0.5\textwidth]{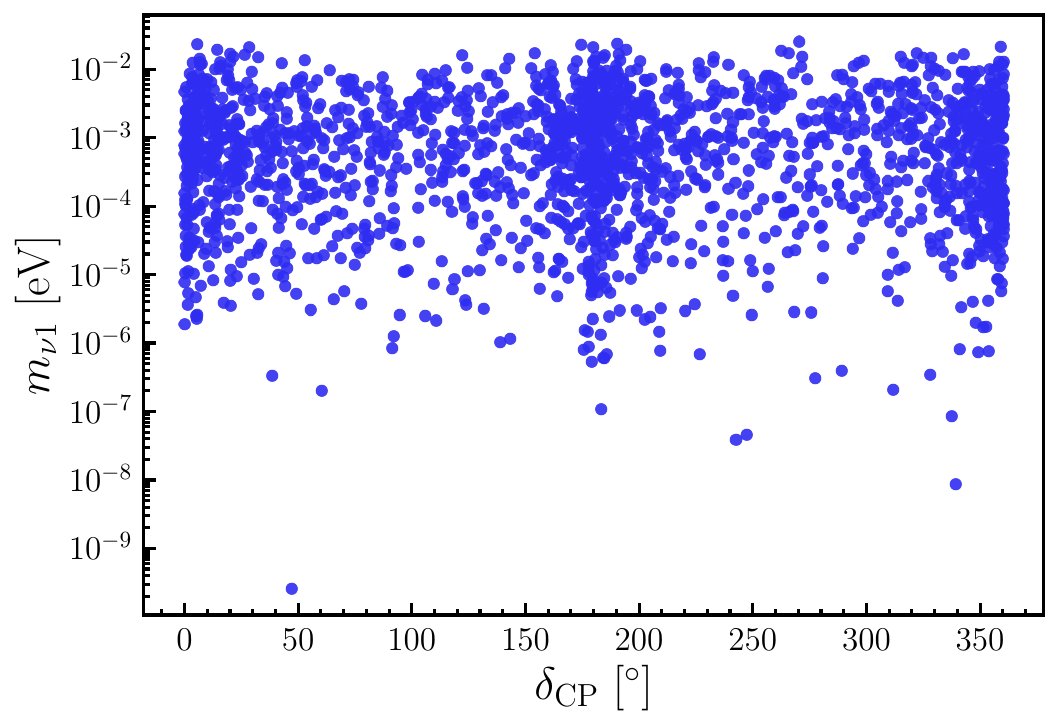} 
\end{center}
\caption{Left panel : Allowed values of the real and imaginary parts of the modular field $\tau$. The region within the magenta line corresponds to the fundamental domain of the modular group. Right panel : Correlation of the lightest neutrino mass $m_1$ to 
the leptonic Dirac CP violating phase $\delta_{\rm CP}$.}
\label{figtau}
\end{figure}

The left panel of Fig.(\ref{figtau}) shows the allowed values for the real and imaginary parts of the modulus field $\tau$. We take all the parameter points satisfying $\chi^2 \leq 9$.

An important result that we have obtained is that the model does not allow inverted ordering (IO) for active light neutrino masses even if we take all the Yukawa couplings to be complex. Thus, all the results given in this paper are for the normal ordering (NO) of active light neutrino masses. Moreover, since all leptonic Yukawa couplings are taken to be real, the only source of complex phases is the imaginary part of $\tau$.

The results are shown in Fig.~(\ref{figtau}). The left panel of this figure shows the allowed values of the real and imaginary parts of the modular field $\tau$. It can be seen that the majority of the values of $\tau$ within the fundamental domain are allowed by the constraints on lepton masses and mixing. We also examined the correlation between the lightest neutrino mass $m_1$ and the Dirac CP phase $\delta_{\text{CP}}$, shown in the right panel of Fig.~(\ref{figtau}). Viable solutions span the full range of $\delta_{\rm CP}$, with a higher density of points near the CP-conserving regions $\delta_{\rm CP}\simeq0^\circ,180^\circ,360^\circ$. Most viable solutions lying roughly in the meV to $10^{-2}$ eV region.

%%%%%%%%%%%%%%%%%%%%%%%%%%%%%%%%%

\section{Charged lepton flavour violation}
\label{sec:lfv}

\begin{table}[t]
\centering
\renewcommand{\arraystretch}{1.2}
\setlength{\tabcolsep}{8pt}
\begin{tabular}{lcc}
\toprule[0.2mm]
\hline
\textbf{Process} & \textbf{Current bound} & \textbf{Projected sensitivity} \\
\midrule[0.1mm]
\hline
$\mu \to e \gamma$ 
& $\operatorname{Br}<1.5 \times 10^{-13}$ 
& $\operatorname{Br}\sim 6 \times 10^{-14}$ \\

$\tau \to e \gamma$ 
& $\operatorname{Br}<3.3 \times 10^{-8}$ 
& $\operatorname{Br}\sim3.3 \times 10^{-9}$ \\

$\tau \to \mu \gamma$ 
& $\operatorname{Br}<4.4 \times 10^{-8}$ 
& $\operatorname{Br}\sim 4.4 \times 10^{-9}$ \\
\bottomrule[0.2mm]
\end{tabular}
\caption{Current bounds and projected sensitivities for the radiative charged lepton flavour violating decays considered in this work. The $\mu\to e\gamma$ projected sensitivity corresponds to the expected MEG II full-run sensitivity \cite{MEGII:2025gzr}, while the $\tau\to e\gamma$ and $\tau\to\mu\gamma$ projected sensitivities correspond to the projected superKEKB/Belle II reach \cite{Hayasaka:2012pj}.}
\label{tab:lfv_exp}
\end{table}

Charged lepton flavour violation provides one of the most sensitive probes of physics beyond the SM. In the SM, charged LFV decay rates are suppressed to completely negligible levels, so any observable signal would constitute an unambiguous signature of new physics. For this reason, processes like $\mu\to e\gamma$ play a central role in current and near-future experimental searches, with $\mu\to e\gamma$ currently providing one of the most stringent bounds and with very promising prospects in the near future. The values for current experimental limits and projected sensitivities for the LFV processes analyzed here can be found in Table~\ref{tab:lfv_exp}.

In our model, LFV arises from the same Yukawa structure responsible for radiative neutrino mass generation. The relevant branching ratios for radiative decays $\ell_\alpha\to\ell_\beta\gamma$ are given by \cite{Toma:2013zsa}
\begin{equation}
\mathrm{Br}(\ell_\alpha \to \ell_\beta \gamma)
=
\mathrm{Br}(\ell_\alpha \to \ell_\beta \nu_\alpha \bar{\nu}_\beta)
\,
\frac{3\,\alpha_{\rm em}}
{64\pi\,G_F^2\,m_{\eta^+}^4}
\left|
\sum_{i=1}^{3}
(x_\nu)_{\beta i}^*
(x_\nu)_{\alpha i}
F\!\left(\frac{m_{N_i}^2}{m_{\eta^+}^2}\right)
\right|^2 ,
\end{equation}
where $m_{\eta^+}$ denotes the charged scalar mass and $m_{N_i}$
$(i=1,2,3)$ are the heavy Majorana neutrino masses. The effective Yukawa coupling in the mass basis entering the LFV amplitudes is given by
\begin{equation}
x_\nu \equiv U_\ell^\dagger \widetilde{Y}_{\nu}=U_\ell^\dagger Y_{\nu}U_N,
\end{equation}

and the loop function is
\begin{equation}
F(x)
=
\frac{1 - 6x + 3x^2 + 2x^3 - 6x^2 \ln x}
{6(1-x)^4}.
\end{equation}

The corresponding one-loop diagrams are omitted here, since they coincide with the standard scotogenic contributions discussed in Ref. \cite{Toma:2013zsa}. For the electromagnetic fine-structure constant and the Fermi constant, we use the values quoted in Ref.~\cite{ParticleDataGroup:2024cfk}
\begin{equation}
\alpha_{\rm em} \equiv \frac{e^2}{4\pi}=1/137.035999177,
\qquad
G_F = 1.16637\times 10^{-5}\,\text{GeV}^{-2}.
\end{equation}

For the specific decays considered in this work, we use the experimental values for SM leptonic decays \cite{ParticleDataGroup:2024cfk}
\begin{align}
&\mathrm{Br}(\mu \to e \nu_\mu \bar{\nu}_e) \simeq 1,\quad \mathrm{Br}(\tau \to e \nu_\tau \bar{\nu}_e) = 0.178,\quad\mathrm{Br}(\tau \to \mu \nu_\tau \bar{\nu}_\mu) = 0.1739.
\end{align}

\begin{figure}[h]
\begin{center}
\includegraphics[width=0.48\textwidth]{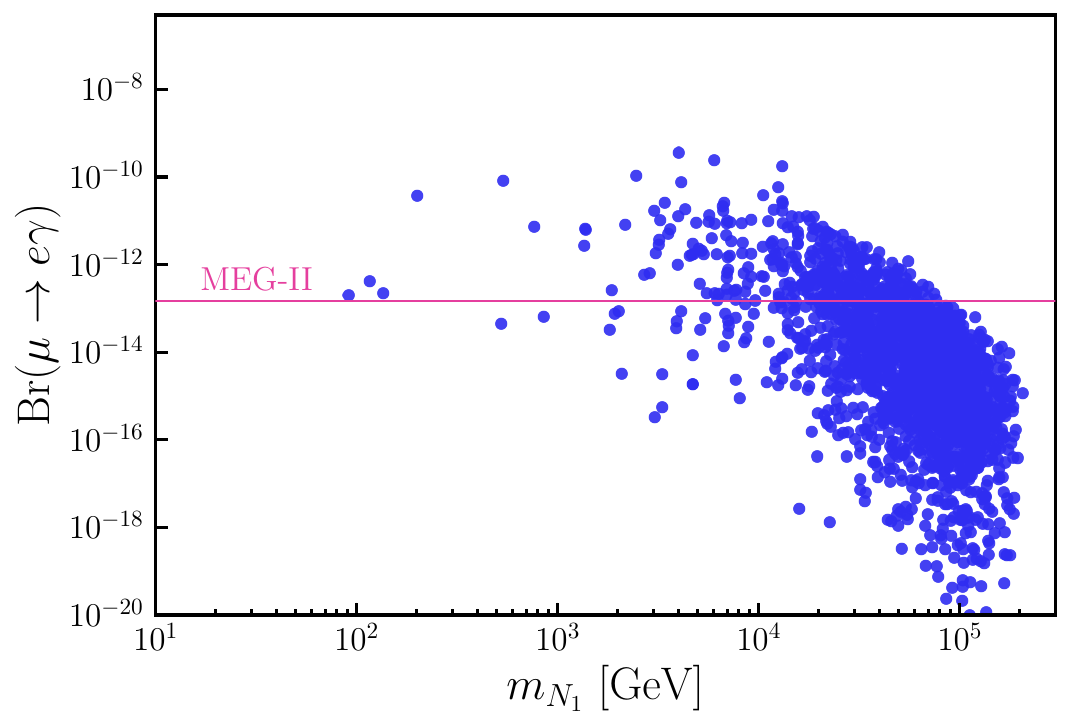} ~~~~~
\includegraphics[width=0.48\textwidth]{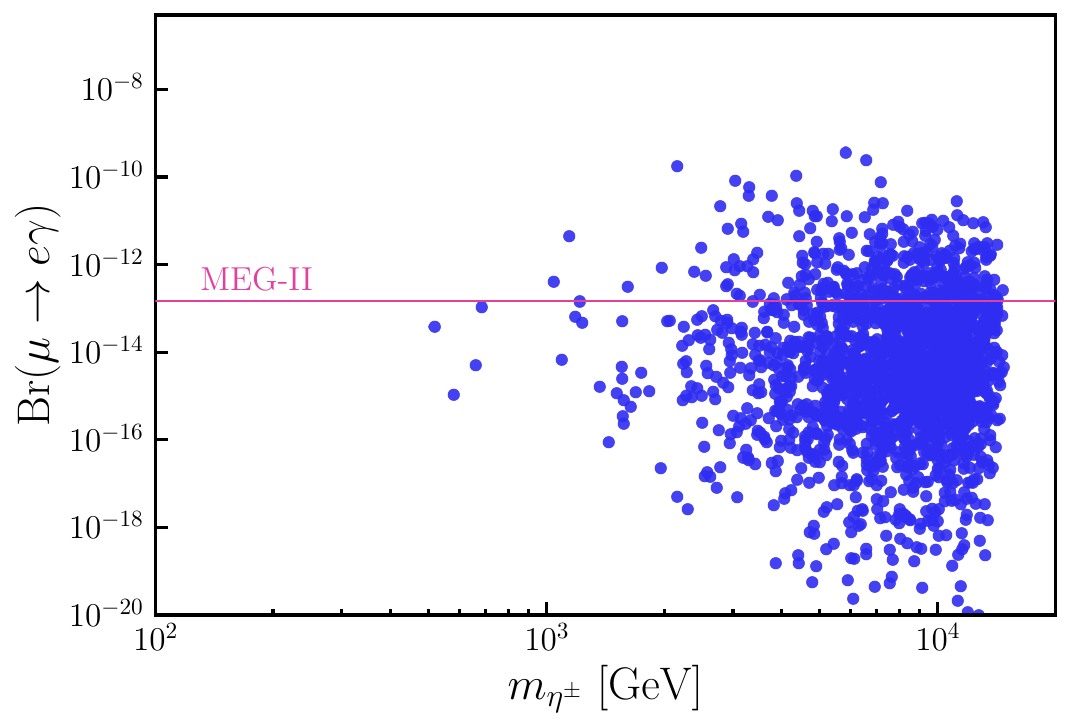} 
\end{center}
\caption{Correlation of $\mathrm{Br}(\mu\to e\gamma)$ with the mass of the lightest heavy neutral lepton $m_{N_1}$ (left panel) and with the charged scalar mass $m_{\eta^\pm}$ (right panel). The horizontal magenta line indicates the current MEG-II sensitivity \cite{MEGII:2025gzr}.}
\label{figLFV1}
\end{figure}

\begin{figure}[h]
\begin{center}
\includegraphics[width=0.48\textwidth]{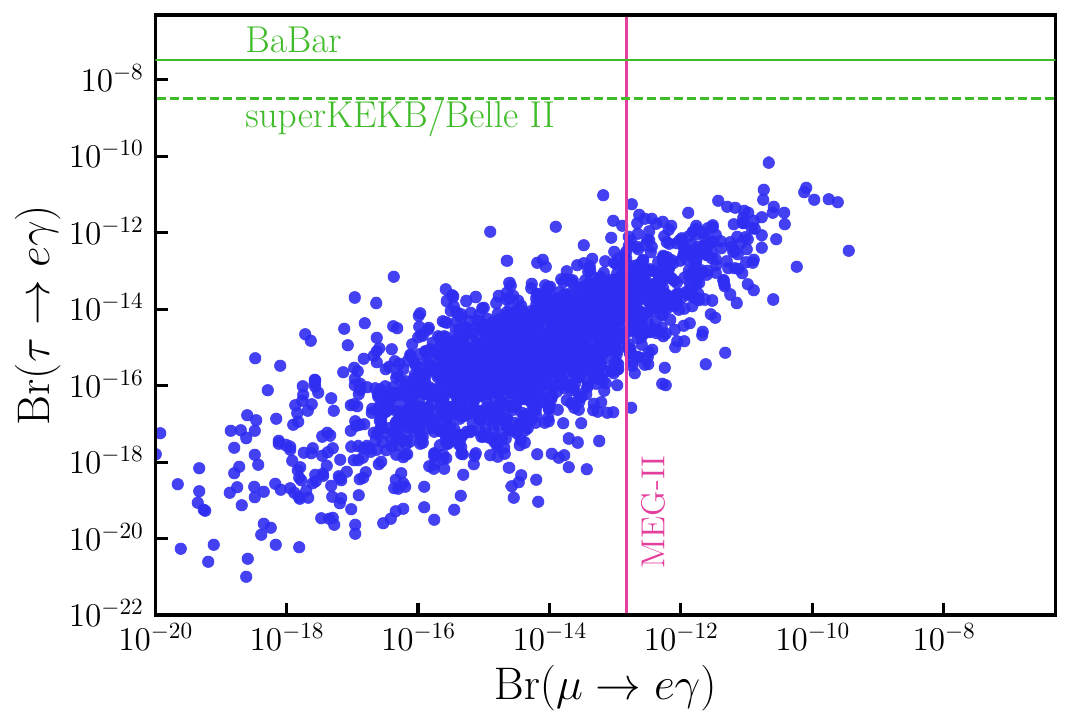} ~~~~~
\includegraphics[width=0.48\textwidth]{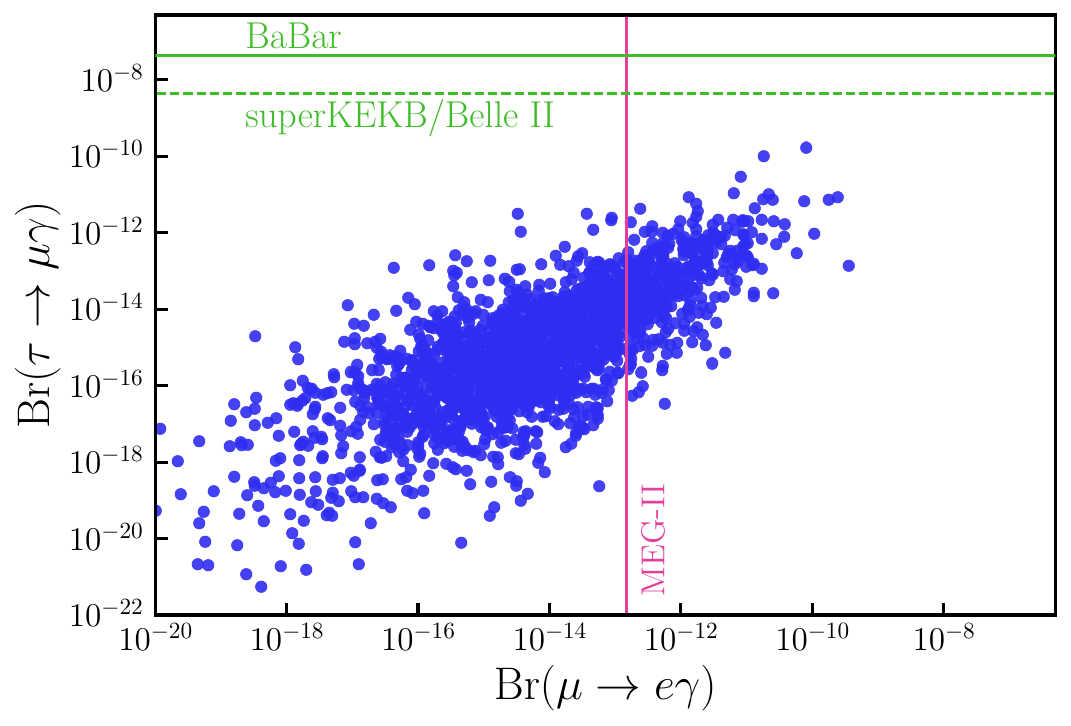} 
\end{center}
\caption{Correlations of Br$(\tau \rightarrow e \gamma)$ and Br$(\tau \rightarrow \mu \gamma)$ against Br$(\mu \rightarrow e \gamma)$.}
\label{figLFV2}
\end{figure}

The numerical results are shown in Fig.~\ref{figLFV1} and Fig.~\ref{figLFV2}. In the left panel of Fig.~\ref{figLFV1}, we show the correlation between $\mathrm{Br}(\mu\to e\gamma)$ and the mass of the lightest heavy neutral lepton $m_{N_1}$. The largest values of $\mathrm{Br}(\mu\to e\gamma)$ are concentrated at lower $m_{N_1}$, while heavier neutral fermions  suppress the branching ratio. The current MEG-II \cite{MEGII:2025gzr} limit excludes a large region of the sub-TeV masses for $m_{N_1}$. The right panel of Fig.~\ref{figLFV1} shows the dependence of $\mathrm{Br}(\mu\to e\gamma)$ on the charged scalar mass $m_{\eta^\pm}$. The correlation is milder than in standard one-loop scotogenic scenarios, since the neutrino mass mechanism strongly correlates the Yukawa couplings with the heavy-neutrino masses, and this dependence dominates the branching ratio.

Fig.~\ref{figLFV2} displays the correlations of $\mathrm{Br}(\tau\to e\gamma)$ and $\mathrm{Br}(\tau\to \mu\gamma)$ with $\mathrm{Br}(\mu\to e\gamma)$. Both panels show a clear positive correlation, reflecting the common radiative origin of the three processes. However, once the $\mu\to e\gamma$ constraint is imposed, the predicted values for the LFV tau decays remain several orders of magnitude below both the current BaBar bounds \cite{BaBar:2009hkt} and the projected SuperKEKB/Belle II sensitivities reported in \cite{Hayasaka:2012pj}. Therefore, in the parameter region explored here, the decay $\mu\to e\gamma$ provides the leading LFV constraint, while the radiative tau decays do not constitute competitive probes.

The LFV predictions become particularly relevant in the $N_1$ fermionic DM scenario as we will show in Section \ref{sec:dark_matter}. The relic abundance is dominated by leptophilic channels mediated by the charged inert scalar. As a consequence, annihilation channels involving the electron flavour are tightly constrained by $\mu\to e\gamma$, whereas the $\mu$--$\tau$ entries can remain sizable and control the dominant annihilation channels.

%%%%%%%%%%%%%%%%%%%%

\section{Dark matter}
\label{sec:dark_matter}

\begin{figure}[h]
\centering
\subfigure[]{\includegraphics[width=.3\linewidth]{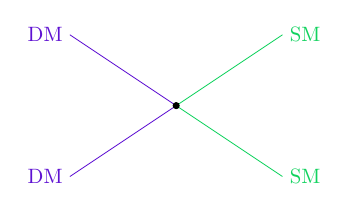}}    
\subfigure[]{\includegraphics[width=.3\linewidth]{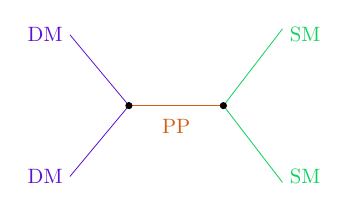}}   \\
\subfigure[]{\includegraphics[width=.3\linewidth]{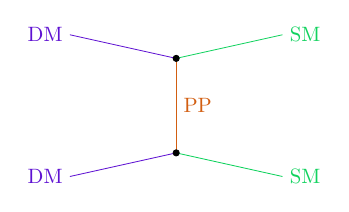}} 
\subfigure[]{\includegraphics[width=.3\linewidth]{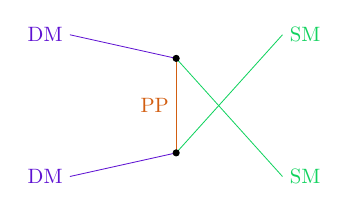}}
\caption{Feynman diagrams for dark matter ($\mathrm{DM}=\chi$, $N_1$, $\Omega_R$) annihilation channels into Standard Model species ($\mathrm{SM}=W^\pm$, $Z$, $h$, $\ell$, $\nu$, $q$) at tree-level. Diagram (a) corresponds to a contact interaction, whereas diagrams (b), (c), and (d) are mediated by a portal particle (PP), which may belong either to the SM or to the dark sector.}
\label{fig:DM_anihilation}
\end{figure}

\begin{figure}[h]
    \centering
    \includegraphics[width=0.3\linewidth]{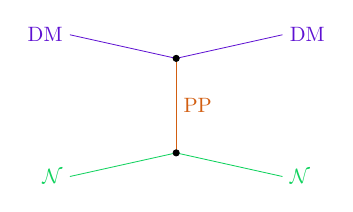}
    \caption{Feynman diagram for elastic scattering between dark matter ($\mathrm{DM}=\chi$, $N_1$, $\Omega_R$) and atomic nucleus at tree-level, which dominates the spin-independent cross section $\sigma_{\mathrm{SI}}$.
}
    \label{fig:DM_DD}
\end{figure}

In this model, the spontaneous breaking of modular symmetry gives rise to a residual discrete $Z_2$ symmetry. The lightest physical states that are odd under this residual symmetry constitute the standard stable dark matter candidates. In this section, we study the phenomenology of these candidates, in particular their potential to account for the dark matter relic density through thermal decoupling in the early Universe, a mechanism commonly known as freeze-out, while confronting the model predictions with experimental constraints from direct detection.

Fields with non-trivial modular weight $k$ inherit a residual charge $Q_{Z_2}=(-1)^k$. According to Table \ref{field_assignments}, the inert scalars $\eta$ and $\varphi$ and the right-handed neutrinos $N_{iR}$ are odd under $Z_2$. This leads to two standard dark matter scenarios: the lightest of the scalar mass eigenstates (scalar dark matter), or the lightest right-handed neutrino $N_R$ (fermionic dark matter). In addition, a non-standard fermionic sub-scenario may exist. The right-handed neutrino $\Omega_R$, which is not stabilized by the residual $Z_2$ symmetry, has only one trilinear interaction with another right-handed neutrino and the field $\sigma$. Consequently, if it is the lightest particle, it becomes metastable with a lifetime exceeding the age of the Universe due to kinematic suppression of its tree-level decay channels.

The DM candidates are assumed to be thermally produced weakly interacting massive particles (WIMP) relics. Assuming that the DM abundance is produced via thermal relic decoupling in the early Universe, it depends on the mass and couplings of the candidate entering the annihilation processes ($2\to 2$). The relic abundance measured by the Planck collaboration $\Omega h^2=0.1200\pm 0.0012$ \cite{Planck:2018vyg} provides the main constraint on the candidate masses, which, for WIMP species with perturbative scalar couplings, are typically of the $\mathrm{TeV}$ scale.

The interaction between particles in the dark and visible sectors opens DM annihilation channels into SM particles after electroweak symmetry breaking; the main tree-level channels are shown in Fig.~\ref{fig:DM_anihilation}. The relative importance of the different annihilation channels determines the phenomenology of each scenario. The same mechanism induces interactions between DM and quarks (see Fig.~\ref{fig:DM_DD}), enabling predictions for direct detection through elastic scattering with nucleons. These predictions must be confronted with the most recent exclusion limits from the LUX-ZEPLIN (LZ) collaboration \cite{LZ:2024zvo} and the projected limits from XENONnT \cite{XENON:2020kmp}, ARGO \cite{Billard:2021uyg}, DarkSide-20k \cite{DarkSide-20k:2017zyg} and DARWIN \cite{Schumann:2015cpa}.

For the analysis of each scenario, we assume a specific mass hierarchy for the dark particle spectrum. We further assume $M_{\mathrm{DM}} < 3.5$ TeV, and the masses of the remaining particles are taken to be sufficiently large to ensure dynamical decoupling from the candidate. Numerical results correspond to a sample of observables obtained from a parameter scan of the scalar potential in Eq.(\ref{eq:scalarpotential}) and the Yukawa couplings in the neutrino sector in Eq.(\ref{eq:nu-N-Om}), demonstrating the viability of the model. Additionally, for each scenario, we provide benchmark points that satisfies the full set of experimental constraints, which will be detailed later, and demonstrates the testability of the model.

The numerical analysis methodology involves implementation of the model in \texttt{FeynRules 2.0} \cite{Alloul:2013bka} and parameter scanning as well as solving Boltzmann equations using the \texttt{micrOMEGAs 6.2.3} code \cite{Alguero:2023zol}. Given the computational cost of the optimization problem, it has been solved via the differential evolution algorithm provided by the \texttt{Python SciPy} package in conjunction with the aforementioned tools.

\subsection{Scalar Dark Matter Scenario}

From Sec.~\ref{sec:scalarsector}, we see that the CP-even/CP-odd mass-eigenstate pairs $S_1$--$P_1$ and $S_2$--$P_2$ are degenerate in mass at tree-level. As a consequence, the scalar dark matter candidate is a complex scalar state. In this work, we restrict our analysis to this scenario, and a brief discussion about its validity will be given in the last part of the analysis. The phenomenology of the candidate depends strongly on its singlet-doublet composition. In turn, the mixture is determined by the hierarchy mass splittings, consistently with Fig.~\ref{fig:scalar_mixture}.

We divide the analysis of this scenario into two parts. First, we study the phenomenology of the limiting cases, namely the pure-doublet and pure-singlet regimes, the main processes governing the relic abundance and direct detection, and whether these limiting cases are viable in this model once the phenomenology of the leptonic sector is taken into account. Second, we explore the regions with non-pure textures, contrasting them with the phenomenology of the limiting cases and identifying the viable ones.

The numerical analysis of the scalar sector can be simplified by using the parametrization in terms of the physical mass splittings of the scalars. The relevant parameters are the mass of the lightest complex scalar, $m_{\chi_1}$, and the splittings $\Delta_0^2$ and $\Delta_\pm^2$. Subdominant couplings entering the annihilation and elastic-scattering processes, namely $\lambda_{2,4,8,9,11}$, are fixed to $10^{-2}$, while $\lambda_{14}$ is fixed to $10^{-3}$.

\subsubsection{Pure cases}

To characterize the phenomenology of the DM candidate in the pure limits, we will discuss in general terms the main annihilation channels, direct detection channels and their behaviour in both regimes. We begin by noting that the exact pure limits are not phenomenologically viable, since the active neutrino mass mechanism is directly controlled by the mass splittings of the inert scalars and therefore on the mixing angle. From Eq.~(\ref{eq:actneutrino}), it is clear that the mass matrix vanishes in the exact pure limits. We therefore define as pure limits phenomenologically viable within this model those mixing regions close to the exact pure limits. In particular, we define the $\chi_1$ singlet-like limit by $\sin\beta<0.001$ and the doublet-like limit by $\sin\beta>0.999$, with $\beta\in(0,\pi/2)$.

Assuming a freeze-out scenario, the dark matter relic density is governed by the main dark matter annihilation channels, $(\chi\chi\to\mathrm{SM\,SM})$, where $\mathrm{SM}=h,Z,W^\pm,\ell,\nu,q$, through the diagrams shown in Fig.~\ref{fig:DM_anihilation}. The trilinear Higgs-portal coupling plays a central role. Expanding the potential in Eq.~(\ref{eq:scalarpotential}), one obtains the explicit form of the coupling,
\begin{equation}
\label{eq:higgs_portal}
\lambda_{\chi h}\approx (\lambda_5+\lambda_6)v\sin^2\beta+\lambda_{12}v_\sigma\cos\beta\sin\beta+\lambda_7v\cos^2\beta.
\end{equation}
where we have assumed a scenario very close to the alignment limit ($\alpha\sim 0$), in accordance with the Higgs-physics constraints discussed in Sec.~\ref{sec:scalarsector}.

Annihilation into electroweak gauge bosons typically dominates the relic abundance for a dark matter candidate with a non-negligible inert-doublet component, as occurs in the mostly doublet-like region of $\chi_1$. In this model, the doublet component of the lightest scalar is proportional to $\sin\beta$. Therefore, its interaction with gauge bosons vanishes in the singlet-like limit $\beta\to 0$ and becomes maximal in the doublet-like limit $\beta\to\pi/2$.

For the channels $\chi_1\chi_1\to W^+W^-,ZZ$, diagram (b) is mediated by a Higgs boson and is proportional to the Higgs-portal coupling $\lambda_{\chi h}$. By contrast, diagrams (a), (c), and (d) are controlled by the doublet mixture of $\chi_1$. For the $W^+W^-$ final state, the particles mediating diagrams (c) and (d) are the charged inert scalars $\eta^\pm$, whereas for the $ZZ$ final state the mediators are the neutral inert scalars $\chi_{1,2}$.

In the doublet-like limit, annihilation into gauge bosons is dominated by electroweak interactions. In the $ZZ$ channel, there is a strong cancellation between diagram (a) and the $t$- and $u$-channel contributions when the mediator is the DM candidate itself or a particle nearly degenerate with it. In the $W^+W^-$ channel, this cancellation is only partial because of the mass splitting between the neutral and charged components, which generally enhances the relevance of this channel whenever $\Delta_\pm^2$ is sizable. In the singlet-like limit, the inert--gauge interaction vanishes and the annihilation proceed through diagram (b), mediated by the Higgs boson. This channel therefore becomes relevant in regions close to the $s$-channel pole, $2m_{\chi_1}\sim m_h$, or in regions where the Higgs-portal coupling $\lambda_{\chi h}\approx \lambda_7v$ is sufficiently large.

Annihilation into pairs of Higgs bosons, $\chi_1\chi_1\to hh$, can also become relevant because of the presence of trilinear scalar couplings, which may dominate in some regimes, and because this channel is less sensible at the different DM candidate mixtures except through the angular dependence entering the Higgs-portal coupling. In this channel, diagram (a) is proportional to the four-point coupling between $\chi_1$ and the Higgs, whereas diagrams (c) and (d) are proportional to the Higgs-portal coupling and are mediated by the inert scalars $\chi_{1,2}$. Diagram (b) depends both on the Higgs-portal coupling and on the Higgs trilinear coupling, and is mediated by a Higgs boson.

As mentioned above, this channel does not discriminate among mixtures of the DM candidate if $\lambda_{5,6,7}$ are taken to be comparable. However, in the singlet-like limit it becomes especially relevant because of the reduced importance of annihilation into gauge bosons. For DM masses above $100$ GeV, the process lies away from the pole in diagram (b), while diagrams (c) and (d) are suppressed by the large masses of the inert scalars. One therefore expects diagrams (a) and (b) to provide the dominant contributions for the singlet-like limit.

Finally, annihilation into SM fermions ($\ell$, $q$, $\nu$) through diagram (b), mediated by a Higgs boson is typically subdominant compared with annihilation into Higgs pairs since the Yukawa coupling is proportional to the fermion mass. In our model, however, there are two additional sources contributing to this channel arising from the doublet component of $\chi_1$. The coupling to the $Z$ boson allows for an additional $s-$channel mediated by the $Z$, which reaches its maximum relevance in the doublet-like limit. On the other hand, annihilation into active neutrinos receives, in addition to the previous contribution, contributions from diagrams (c) and (d), mediated by the heavy neutrinos $N_R$ and proportional to the Yukawa couplings $\tilde Y_D$ that can be specially relevant for regions with large Yukawa couplings. In the doublet-like limit, annihilation into fermions becomes more relevant because of the presence of these additional contributions.

Elastic scattering between the DM candidate and atomic nuclei (denoted by $\mathcal{N}$), $\chi\mathcal N\to \chi\mathcal N$, proceeds through the diagram shown in Fig.~\ref{fig:DM_DD}. In the singlet-like limit, this process is mediated by a Higgs boson and the corresponding cross section is controlled by the Higgs-portal coupling, whereas in the doublet-like limit the doublet component provides an electroweak contribution mediated by the $Z$ boson.

In summary, in the singlet-like limit a larger DM abundance is expected because of the suppression of annihilation into gauge bosons, and in general both the relic abundance and elastic scattering with nucleons are dominated by the Higgs-portal coupling and resonant effects. In the doublet-like limit, annihilation is more efficient because of the opening of annihilation channels into gauge bosons and, consequently, a smaller abundance is expected, while elastic scattering with nucleons is enhanced by electroweak interactions mediated by the $Z$ boson. These two limiting behaviours will serve as the qualitative reference for interpreting the phenomenology of the mixed scenarios.

\subsubsection{Mixed case and numerical results}

\begin{table}[t]
\centering
\begin{tabular}{c c c c c c c}
\toprule[0.2mm]
\hline
BP & $m_{\chi_1}$ [GeV] & $m_{\chi_2}$ [GeV] & $m_{\eta^\pm}$ [GeV] & $\beta/\pi$ & $\Omega h^2$ & $\sigma_{\rm SI}$ [cm$^2$] \\
\hline
BP1 & $147.2$ & $1635.7$ & $1631.1$ & $0.002$ & $0.1205$ & $4.29\times10^{-48}$ \\

BP2 & $2670.6$ & $4125.7$ & $3680.4$ & $0.253$ & $0.1192$ & $1.59\times 10^{-51}$ \\

BP3 & $2673.1$ & $11126.5$ & $3382.5$ & $0.472$ & $2.57\times 10^{-4}$ & $5.99\times 10^{-48}$ \\
\bottomrule[0.2mm]
\end{tabular}
\caption{Benchmark points for the scalar DM scenario. BP1 corresponds to a point lying below the current LZ bound and within the projected reach of XENONnT but it saturates the relic abundance. BP2 represents a mixed scenario close to maximal singlet-doublet mixture, compatible with relic abundance and direct detection bounds, but beyond the projected XENONnT sensitivity. BP3 corresponds to a viable mostly doublet-like scenario, it accounts only for a subdominant fraction of the observed dark matter abundance.}
\label{tab:scalar_BPs}
\end{table}

\begin{figure}
    \centering
    \subfigure[]
    {\includegraphics[width=0.49\linewidth]{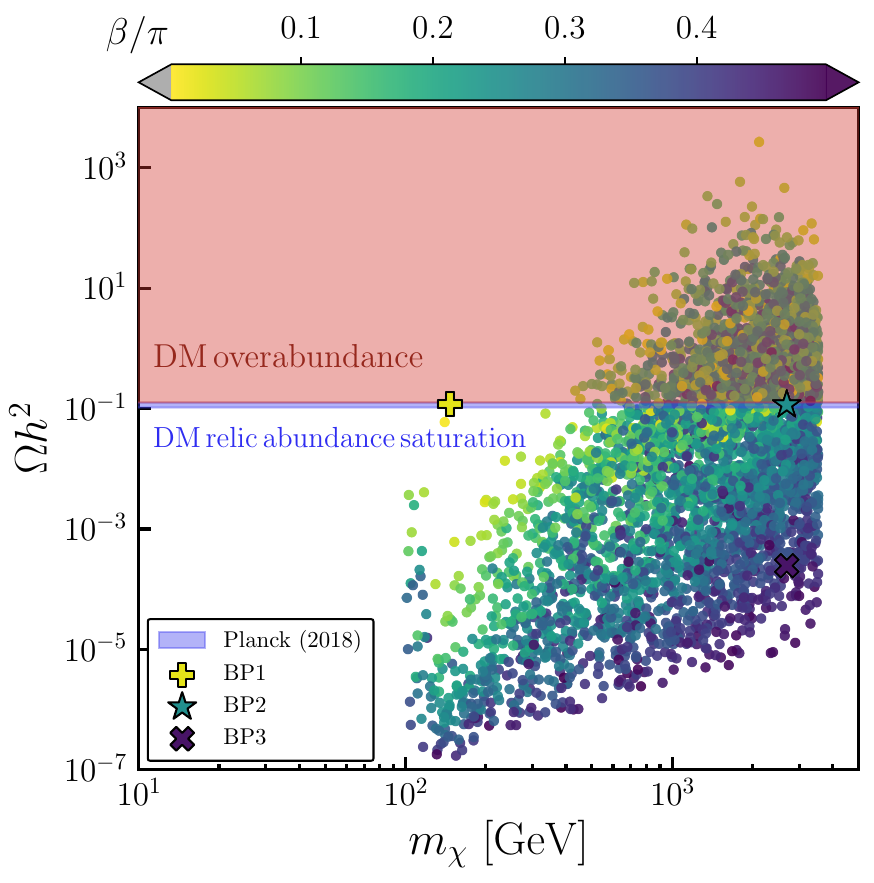}}
    \subfigure[]
    {\includegraphics[width=0.5\linewidth]{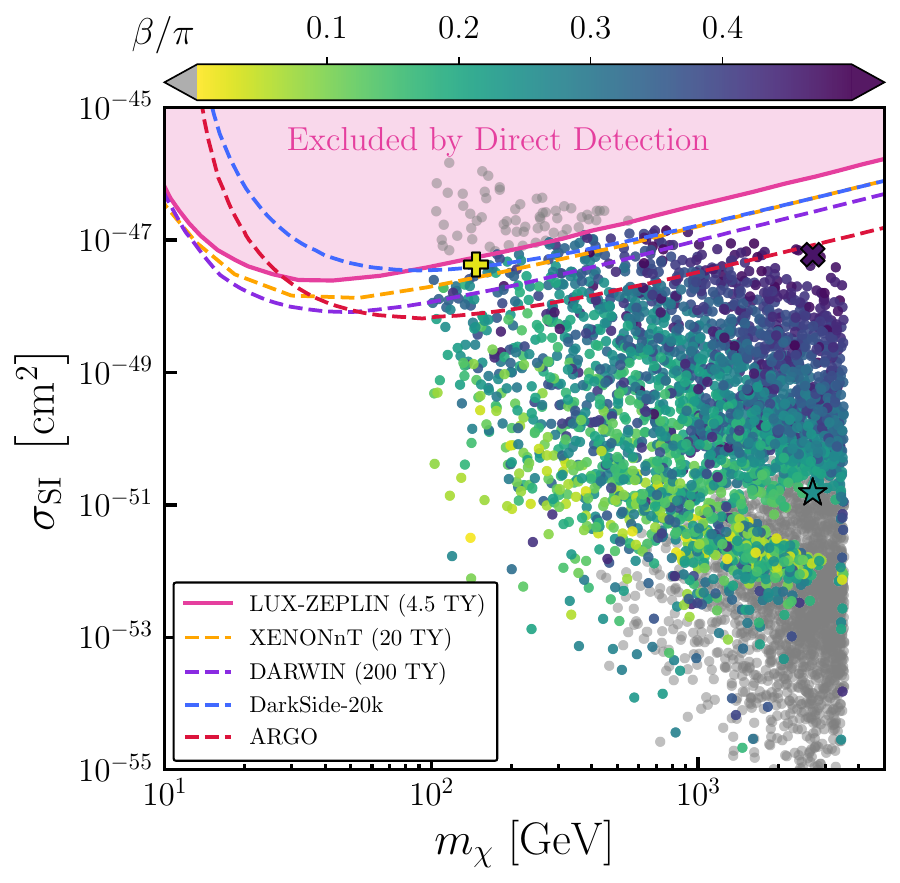}}
    \caption{Dependence of DM relic density and DM-nucleon elastic scattering cross section on the lightest inert-scalar mass $m_{\chi_1}$ (x-axis) and the mixing angle $\beta/\pi$ (shown by the color scale). Panel (a) shows the relic abundance $\Omega h^2$ where the horizontal dashed line corresponds to the observed DM abundance reported by Planck \cite{Planck:2018vyg}. Panel (b) shows the spin-independent DM-nucleon elastic scattering cross section $\sigma_{\rm SI}$, together with the current LZ exclusion limit \cite{LZ:2024zvo} and the projected next direct detection experiments sensitivity \cite{XENON:2020kmp,Billard:2021uyg,DarkSide-20k:2017zyg,Schumann:2015cpa}. In panel (b), the grey points correspond to overabundant scenarios, $\Omega h^2>\Omega_{\rm obs}h^2$. The benchmark points BP1, BP2 and BP3 illustrate relic compatible and experimentally viable scenarios. BP1 is relic-saturating and experimentally testable. BP2 reproduces the observed relic abundance but is beyond the projected direct detection proyects reach and BP3 is a viable mostly doublet-like configuration with subdominant relic abundance.}
    \label{fig:ScalarDM}
\end{figure}

The richest scalar DM phenomenology is found in the intermediate mixing region between doublet and singlet components, where both the scalar portal and gauge interactions can be simultaneously active. This mixed region interpolates between the phenomenology of the pure limits, opening viable regions compatible with both relic abundance and direct detection constraints.

The parameter scan covers the full range of mixtures considered in our scan, fixing the heavy neutrino masses above $5$ TeV and the visible scalar masses above $7$ TeV in order to suppress the contribution of heavy-neutrino-mediated annihilation channels and to remain away from the $H$ resonance region. In this way, we isolate the role of mixing in the phenomenology of the candidate.

In Fig.~\ref{fig:ScalarDM}, we show the behaviour of the relic abundance and the spin-independent cross section as functions of the candidate mass and mixing angle. The figure shows that, on average, the abundance tends to increase with the mass of $\chi_1$, while scattering with nucleons shows the opposite behaviour. Fig.~\ref{fig:ScalarDM}(a) displays a clear dependence of the abundance on the mixing angle, allowing the pure limits to be identified at the extreme values. The mostly singlet-like region tends to produce larger abundances, whereas the mostly doublet-like region is predominantly underabundant. The intermediate-mixing region naturally populates the region where the observed relic abundance is saturated. Fig. \ref{fig:ScalarDM}(b) shows the value of the elastic DM--nucleon scattering cross section. Here, the grey points below the LZ exclusion line correspond to those yielding an overabundant DM relic density, excluding a large part of the mostly singlet-like region at masses of order TeV. Scattering with nucleons also shows a strong dependence on the mixing angle, the mostly doublet-like region yields larger scattering rates than the mostly singlet-like one, often by several orders of magnitude, leaving only a small surviving portion of the sub-TeV mass region. Most of the intermediate-mixing region is consistent with the LZ bounds, and within the reach of forthcoming direct-detection experiments. It can also be seen that, although there is a strong dependence on the mixture of the candidate, the dispersion of values at fixed mass reflects an additional dependence on both the Higgs-portal coupling and the mass splittings.

To illustrate the viability of $\chi_1$ as a DM candidate, we select the representative benchmark points shown in Table \ref{tab:scalar_BPs}, each of which illustrates distinct phenomenological behaviours. In BP1, $\chi_1$ reproduces the observed DM relic abundance, and its scattering with nucleons is sufficiently small to evade the current LZ direct detection bound, while still lying within the projected reach of XENONnT. This benchmark therefore represents a testable scenario of the model. In BP2, $\chi_1$ also reproduces the observed DM relic abundance and is compatible with direct detection bounds, but it is not expected to be probed by near-future direct detection experiments because of its suppressed interaction with nucleons. This benchmark corresponds to the region close to maximal singlet-doublet mixing. In BP3, the DM candidate satisfies $\Omega h^2<\Omega_{\text{obs}}h^2$ and is compatible with direct detection, but it does not reproduce the observed abundance. This benchmark represents a scenario near the doublet-like regime. Taken together, these three benchmarks summarize the main viable phenomenological regimes of the scalar dark matter scenario.

\subsubsection{The role of the radiative splitting}

The distinctive feature of this model is that at tree-level, the light inert CP-even and CP-odd states are exactly degenerate but a mass splitting between them is radiatively generated. The relevant question is whether the two states can be dynamically distinguished by the thermal bath at freeze-out. If the splitting remains small compared to the freeze-out temperature, both components stay thermally populated with nearly identical weights and the complex-scalar description remains valid.

Here, the radiative splitting is controlled by the effective coupling $\lambda_{\rm eff}$ defined in Eq.~\eqref{eq:lambdaeff} induced after symmetry breaking. Using Eq.~\eqref{eq:radsplit}, the splitting can be estimated as
\begin{equation}
\begin{aligned}
\Delta m_1 \equiv |m_{P_1}-m_{S_1}|
\simeq
\frac{|\delta| \cos^2\beta}{m_{\chi_1}},
\end{aligned}
\end{equation}
where $\delta=\lambda_\text{eff}v_\sigma^2/2$, so
\begin{equation}
\begin{aligned}
\Delta m_1
\simeq
\frac{|\lambda_{\rm eff}|\,v_\sigma^2\cos^2\beta}{2m_{\chi_1}}.
\end{aligned}
\end{equation}
The relevant thermal scale is the freeze-out temperature $T_f \simeq \frac{m_{\chi_1}}{x_f}$, so that the condition for thermal indistinguishability is simply
\begin{equation}
\begin{aligned}
\Delta m_1 \ll T_f.
\end{aligned}
\end{equation}

\begin{table}[t]
\centering
\begin{tabular}{c c c c c c c}
\toprule[0.2mm]
\hline
BP & $v_\sigma$ [GeV] & $x_f$ & $\lambda_{\rm eff}$ & $\Delta m_1$ [GeV] & $T_f$ [GeV] & $\Delta m_1/T_f$ \\
\hline
BP1 & $1.999\times 10^{4}$ & $23.6$ & $-4.47\times 10^{-7}$ & $6.07\times 10^{-1}$ & $6.24$ & $9.72\times 10^{-2}$ \\

BP2 & $1.880\times 10^{4}$ & $26.8$ & $5.82\times 10^{-10}$ & $1.89\times 10^{-5}$ & $99.7$ & $1.90\times 10^{-7}$ \\

BP3 & $1.433\times 10^{4}$ & $33.0$ & $-2.03\times 10^{-7}$ & $6.01\times 10^{-5}$ & $81.0$ & $7.42\times 10^{-7}$ \\
\bottomrule[0.2mm]
\end{tabular}
\caption{Thermal quantities of scalar DM benchmark points. These quantities are relevant to understand the implications of loop-induced CP-even/CP-odd mass splitting on the scalar DM dynamics.}
\label{tab:scalar_split}
\end{table}

For the benchmark points in Table~\ref{tab:scalar_BPs}, the values of relevant parameters are shown in Table~\ref{tab:scalar_split}, where $x_f$ is computed with \texttt{micrOMEGAS 6.2.3}. This indicates that the radiatively induced splitting is always smaller than the freeze-out temperature, and in BP2 and BP3 it is many orders of magnitude smaller. Therefore, the thermal bath does not have enough resolution to distinguish the CP-even and CP-odd components as separate species. This is precisely the situation realized by our benchmark points.

\subsection{Fermionic Dark Matter}

\begin{table}[t]
\centering
\begin{tabular}{c c c c c c c c}
\toprule[0.2mm]
\hline
BP & $m_{N_1}$ [GeV] & $m_{\Omega_R}$ [GeV] & $m_{\eta^\pm}$ [GeV] & $\Omega h^2$ & $\sigma_{\rm SI}$ [cm$^2$] & Br$(\mu\to e\gamma)$ & Br$(\tau\to \mu\gamma)$ \\
\hline
BP1 & $1028.5$ & $9658.1$ & $6041.9$ & $0.1201$ & $2.77\times 10^{-51}$ & $1.23\times 10^{-13}$ & $9.59\times 10^{-11}$ \\

BP2 & $7017.4$ & $139.9$ & $5321.5$ & $0.123$ & $2.82\times 10^{-53}$ & $3.6\times 10^{-14}$ & $5.44\times 10^{-13}$ \\
\bottomrule[0.2mm]
\end{tabular}
\caption{Benchmark points for the fermionic dark matter scenario. In BP1 the lightest $Z_2$-odd fermion $N_1$ is the dark matter candidate, whereas BP2 corresponds to the non-standard scenario in which the metastable state $\Omega_R$ plays the role of dark matter.}
\label{tab:fermion_BPs}
\end{table}

\begin{figure}
    \centering
    \subfigure[]{\includegraphics[width=0.485\linewidth]{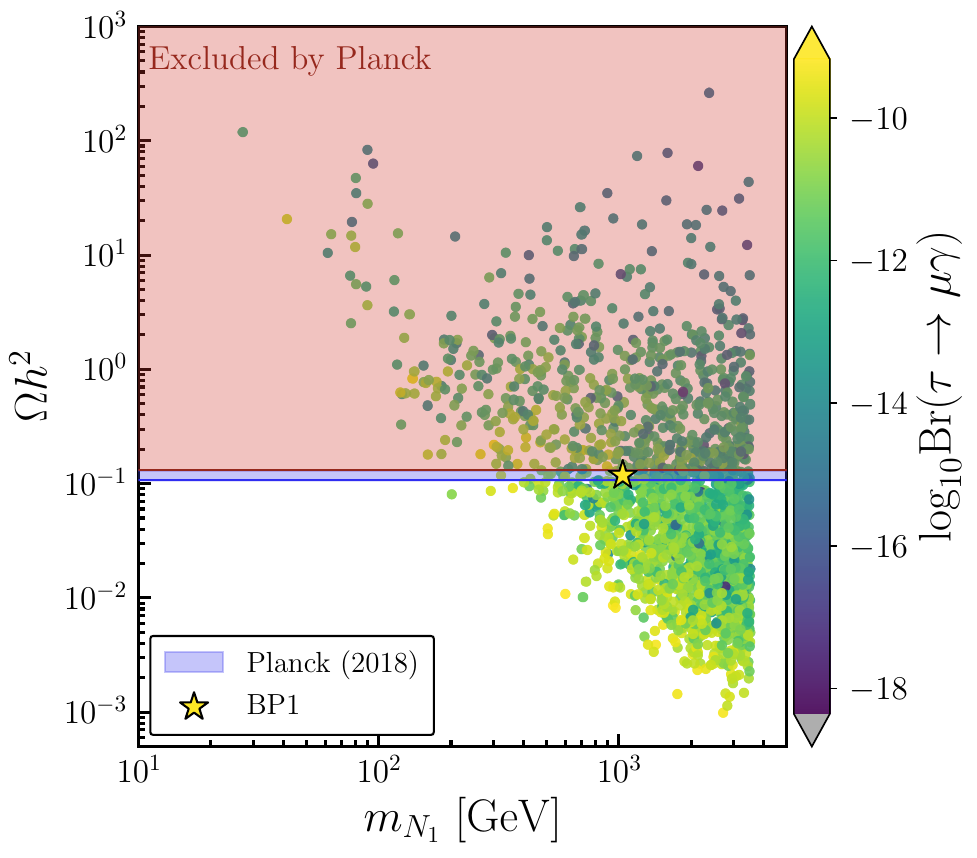}
    }
    \subfigure[]{\includegraphics[width=0.49\linewidth]{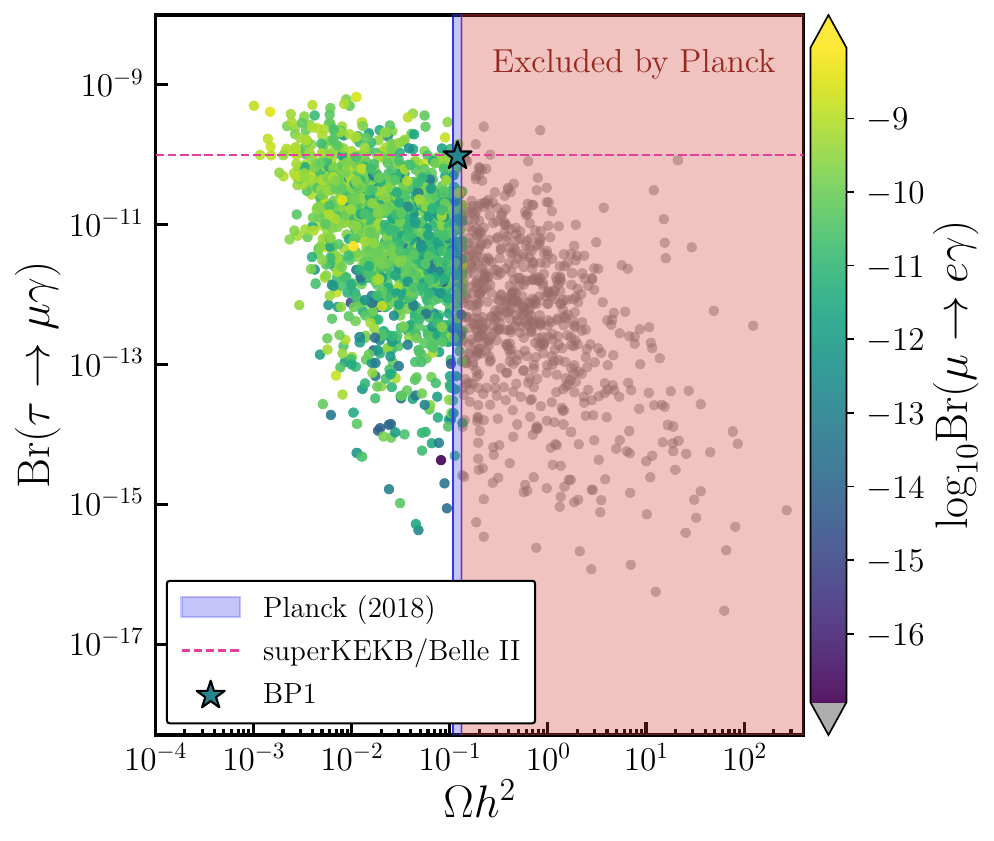}
    }
    \caption{Left panel: relic abundance $\Omega h^2$ as a function of the fermionic dark matter mass $m_{N_1}$, with the color code indicating $\mathrm{Br}(\tau\to\mu\gamma)$. Right panel: correlation between $\Omega h^2$ and $\mathrm{Br}(\tau\to\mu\gamma)$, with the color code showing $\mathrm{Br}(\mu\to e\gamma)$. The observed relic abundance is achieved in the region where leptophilic annihilation of $N_1$ is sufficiently efficient, while the electronic Yukawa entries remain suppressed enough to satisfy the current $\mu\to e\gamma$ bound.}
    \label{fig:Fermion_DM}
\end{figure}

A second viable DM realization of the model arises when the lightest $Z_2$-odd fermion is the right-handed neutrino $N_1$. In contrast to the scalar case, the scalar portal is controlled by the $CP$-even mixing angle $\alpha$, so near the alignment limit the relic abundance is not mainly controlled by the Higgs-portal in Eq.~\eqref{eq:higgs_portal}, but by the Yukawa interactions responsible for the radiative neutrino mass mechanism.

Numerical analysis shows that the dominant annihilation channel is the process $N_1N_1 \to \ell^+\ell^-$ through $t$- and $u$-channel diagrams mediated by the charged inert scalar, which induce leptophilic annihilation into charged leptons while the $s-$channel is suppressed by the Higgs alignment limit as already discussed. In the viable benchmark region we find that the main final states are $\mu$ and $\tau$ leptons, so the relic density is largely controlled by the Yukawa entries involving the $\mu$ and $\tau$ flavours, together with the charged-scalar mass.
This has an immediate phenomenological consequence: the same Yukawa structure that fixes the relic abundance also induces charged-lepton flavour violation. In particular, the process $\tau\to\mu\gamma$ is correlated with the DM abundance, while $\mu\to e\gamma$ acts as a constraint on the electronic Yukawa entries. In this sense, the viable parameter space is shaped more efficiently by LFV constraints than by direct detection. Indeed, the spin-independent DM--nucleon cross section, mediated by the Higgs-portal, is typically far below the current sensitivity of direct detection experiments, while $\mu\to e\gamma$ and $\tau\to\mu\gamma$ may lie close to the present experimental limits.

The numerical results of our analysis can be found in Fig.~\,\ref{fig:Fermion_DM}. Panel (a) shows the correlation between relic abundance and LFV observables. Here, the relic abundance is presented as a function of $m_{N_1}$ with a color code for $\mathrm{Br}(\tau\to\mu\gamma)$. Panel (b) shows the correlation between $\Omega h^2$ and $\mathrm{Br}(\tau\to\mu\gamma)$ using $\mathrm{Br}(\mu\to e\gamma)$ as color code. In this way, one can directly see that the observed relic abundance is achieved in the region where leptophilic annihilation is efficient enough, while the electronic flavour entries remain sufficiently suppressed to evade the current $\mu\to e\gamma$ bound.

A representative benchmark point for this scenario is included in Table~\ref{tab:fermion_BPs} as BP1, where $N_1$ reproduces the observed relic abundance through annihilation channels dominated by the $\mu$--$\tau$ sector and remains compatible with current LFV bounds. Indeed, the value of $\mathrm{Br}(\mu\to e\gamma)$ predicted by BP1 lies within the projected MEG II sensitivity \cite{MEGII:2025gzr}.

\begin{figure}
    \centering
    \includegraphics[width=0.55\linewidth]{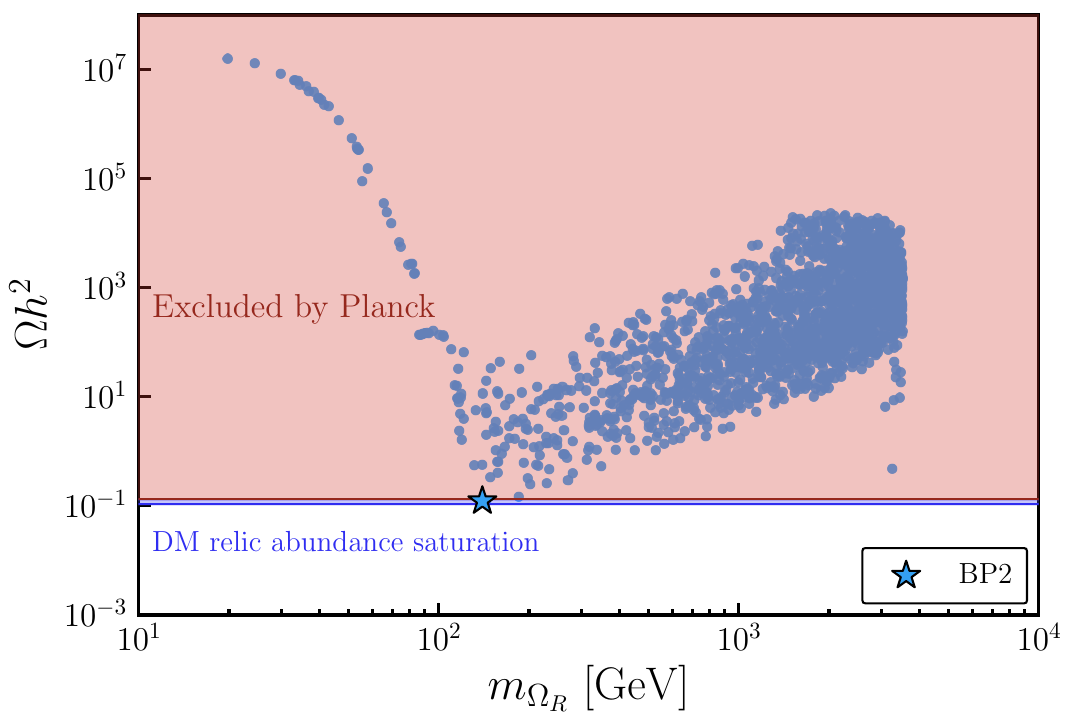}
    \caption{Relic abundance $\Omega h^2$ as a function of the fermionic dark matter mass $m_{\Omega_R}$ in the metastable $\Omega_R$ scenario. The observed Planck value is reached only in the narrow region where annihilation into Higgs bosons is efficient.}
    \label{fig:Omega_DM}
\end{figure}

A second fermionic realization of DM is obtained when $\Omega_R$ is the lightest state in the fermionic dark sector. In contrast to the $N_1$ scenario, $\Omega_R$ is not stabilized by the remnant $Z_2$ symmetry. However, when it is lighter than the remaining fermionic states and the relevant scalar-mediated decay channels are kinematically closed, it becomes effectively metastable on cosmological timescales and can therefore play the role of DM.

The phenomenology of this scenario differs qualitatively from that of $N_1$ DM. Here, the only relevant annihilation channel is annihilation into Higgs bosons, $\Omega_R\Omega_R\to hh$ through $t$- and $u$-channel exchange of $\Omega_R$ itself. The corresponding amplitudes are controlled by the coupling of $\Omega_R$ to the Higgs sector, which is proportional to $m_{\Omega_R}$, and not by the Yukawa interaction with leptons. As a consequence, the viable region is concentrated in the low-mass Higgs-dominated regime, while for values of $m_{\Omega_R}$ away from this region the annihilation cross section becomes inefficient and the relic abundance rapidly grows above the observed value.

This behaviour is illustrated in Fig.~\ref{fig:Omega_DM}, where the relic abundance is shown as a function of $m_{\Omega_R}$. The scan indicates that only a narrow region near the threshold for annihilation into Higgs bosons can reproduce the observed relic abundance, whereas heavier $\Omega_R$ masses typically lead to overabundance.

Regarding direct detection, we can conclude the same behaviour as in the $N_1$ case. Since the interaction of $\Omega_R$ with quarks is mediated by the Higgs sector, it is suppressed by the alignment limit and the large singlet scale $v_\sigma$, resulting in an extremely small spin-independent DM--nucleon cross section.

A viable benchmark point BP2 for this scenario is included in Table~\ref{tab:fermion_BPs}, where $\Omega_R$ reproduces the observed relic abundance, remains compatible with current LFV bounds, and predicts a negligible direct detection signal.

%%%%%%%%%%%%%%%%%%%%

\section{Summary and conclusions}
\label{sec:summary}

We have proposed a two-loop radiative neutrino mass model based on the finite modular flavour symmetry $\Gamma_4\simeq S_4$ supplemented by a discrete $Z_3$ symmetry. After the spontaneous breaking of the $S_4$ modular symmetry, the theory retains a remnant $Z_2$ parity that plays a dual role: it guarantees that the leading contribution to active neutrino masses is radiative, while at the same time stabilizes the lightest $Z_2$-odd state and thereby providing dark matter candidates.

A distinctive feature of the model is that the structure of the scalar sector naturally forbids the one-loop contribution for the active neutrino masses. In particular, the $Z_3$ symmetry forbids the quartic interaction that would induce neutrino masses at one-loop level, while the modular weight assignments prevent the tree-level splitting between the CP-even and CP-odd inert states. This splitting is instead generated radiatively through the effective interaction induced after integrating out the heavy fermions. In this way, the model provides a concrete realization in which the small lepton number violating parameter required in one-loop scotogenic models emerges dynamically.

In the lepton sector, we have shown that the model can reproduce the observed charged-lepton masses and neutrino oscillation data for normal ordering, while the modular structure leads to non-trivial correlations among neutrino observables and the modulus parameter. The same Yukawa couplings responsible for neutrino mass generation also induce charged lepton flavour violation, where $\mu\to e\gamma$ provides the dominant LFV constraint over the parameter space. By contrast, once this bound is imposed, the predicted rates for $\tau\to e\gamma$ and $\tau\to\mu\gamma$ remain well below current and near-future sensitivities in the region explored in our analysis.

The DM phenomenology of the model is particularly rich. In the scalar scenario, the lightest inert complex scalar $\chi_1$ interpolates between singlet-like and doublet-like regimes through the mixing angle of the inert sector. The pure limits provide useful qualitative benchmarks: the singlet-like regime tends to yield larger relic abundance and smaller direct detection rates, whereas the doublet-like regime enhances annihilation into gauge bosons and typically leads to underabundant DM together with larger elastic scattering rates. The most interesting behaviour arises in the intermediate-mixing region, which qualitatively differs from the phenomenology of conventional pure singlet or pure doublet scenarios. Here both the scalar portal and gauge interactions contribute. We find viable points compatible with the observed relic abundance and present direct detection bounds, including benchmark configurations that remain testable in upcoming experiments.

In the case of fermionic DM, when the lightest $Z_2$-odd fermion is $N_1$, the relic density is not controlled mainly by the Higgs portal, but by leptophilic annihilation channels mediated by the charged inert scalar. This leads to a characteristic interplay between dark matter and flavour observables: the same Yukawa structure that fixes the relic abundance also controls LFV. In this regime, $\mu\to e\gamma$ constrains the annihilation channels dependent on the electron flavour, whereas the $\mu$--$\tau$ sector can remain sizable and dominate the annihilation processes. Therefore, the viable parameter space is constrained more efficiently by LFV than by direct detection, whose rates are typically far below current sensitivity. In addition, we found that a non-standard metastable dark matter scenario with $\Omega_R$ is possible, although only in a rather narrow mass region where annihilation into Higgs bosons is sufficiently efficient.

Overall, our results show that this modular two-loop model provides a coherent and predictive framework in which neutrino masses, lepton flavour violation, and dark matter are closely intertwined. The model allows viable scalar and fermionic dark matter realizations while preserving radiative origin for neutrino masses. From the phenomenological point of view, the most promising probes are the continued improvement in $\mu\to e\gamma$ searches and future direct-detection experiments targeting the mixed scalar region. These complementary searches can test a significant fraction of the viable parameter space.

%%%%%%%%%%%%%%%%%%%%

\section{Acknowledgments}

AECH is supported by ANID-Chile FONDECYT 1261103, 1241855, ANID CCTVal CIA250027, ANID – Millennium Science Initiative Program $ICN2019\_044$ and ICTP through the Associates Programme (2026-2031). J. E-P was supported by UTFSM (Chile) under the Postdoctoral Research Program and partially supported by ANID (Chile) under FONDECYT Grant No. 3260491.
SK is supported by ANID-Chile FONDECYT 1230160, and ANID – Millennium Science Initiative Program $ICN2019\_044$. 
DSA supported by Fondecyt 1250776, and ANID CCTVal CIA250027.

%%%%%%%%%%%%%%%%%%%%%%%%%%%%%
\appendix

%%%%%%%%%%%%%%%%%%%%

\section{The product rules of the $S_{4}$ discrete group}
\label{S4}

The $S_{4}$ is the smallest non abelian group having doublet,
triplet and singlet irreducible representations. $S_{4}$ is the group of
permutations of four objects, which includes five irreducible
representations, i.e., $\mathbf{1,1^{\prime },2,3,3^{\prime }}$ fulfilling
the following tensor product rules \cite{Ishimori:2010au,Penedo:2018nmg} 
 \medskip

\noindent \textbf{$\mathbf{1} \otimes \mathbf{r} = \mathbf{r}$}
\begin{equation}
\mathbf{1} \otimes \mathbf{r}: \qquad (a)_{\mathbf{1}} \otimes (b_i)_{\mathbf{r}} = (a b_i)_{\mathbf{r}}
\end{equation}

\medskip

\noindent \textbf{$\mathbf{1}' \otimes \mathbf{1}' = \mathbf{1}$}
\begin{equation}
\mathbf{1}' \otimes \mathbf{1}': \qquad (a)_{\mathbf{1}'} \otimes (b)_{\mathbf{1}'} = (a b)_{\mathbf{1}}
\end{equation}

\medskip

\noindent \textbf{$\mathbf{1}' \otimes \mathbf{2} = \mathbf{2}$}
\begin{equation}
\mathbf{1}' \otimes \mathbf{2}: \qquad (a)_{\mathbf{1}'} \otimes \begin{pmatrix} b_1 \\ b_2 \end{pmatrix}_{\mathbf{2}} = \begin{pmatrix} a b_1 \\ -a b_2 \end{pmatrix}_{\mathbf{2}}
\end{equation}

\medskip

\noindent \textbf{$\mathbf{1}' \otimes \mathbf{3} = \mathbf{3}'$}
\begin{equation}
\mathbf{1}' \otimes \mathbf{3}: \qquad (a)_{\mathbf{1}'} \otimes \begin{pmatrix} b_1 \\ b_2 \\ b_3 \end{pmatrix}_{\mathbf{3}} = \begin{pmatrix} a b_1 \\ a b_2 \\ a b_3 \end{pmatrix}_{\mathbf{3}'}
\end{equation}

\medskip

\noindent \textbf{$\mathbf{1}' \otimes \mathbf{3}' = \mathbf{3}$}
\begin{equation}
\mathbf{1}' \otimes \mathbf{3}': \qquad (a)_{\mathbf{1}'} \otimes \begin{pmatrix} b_1 \\ b_2 \\ b_3 \end{pmatrix}_{\mathbf{3}'} = \begin{pmatrix} a b_1 \\ a b_2 \\ a b_3 \end{pmatrix}_{\mathbf{3}}
\end{equation}

\medskip

\noindent \textbf{$\mathbf{2} \otimes \mathbf{2} = \mathbf{1} \oplus \mathbf{1}' \oplus \mathbf{2}$}
\begin{align}
\mathbf{1}&: \quad a_1 b_2 + a_2 b_1 \\
\mathbf{1}'&: \quad a_1 b_2 - a_2 b_1 \\
\mathbf{2}&: \quad \begin{pmatrix} a_2 b_2 \\ a_1 b_1 \end{pmatrix}
\end{align}

\medskip

\noindent \textbf{$\mathbf{2} \otimes \mathbf{3} = \mathbf{3} \oplus \mathbf{3}'$}
\begin{align}
\mathbf{3}&: \quad \begin{pmatrix} a_1 b_2 + a_2 b_3 \\ a_1 b_3 + a_2 b_1 \\ a_1 b_1 + a_2 b_2 \end{pmatrix} \\
\mathbf{3}'&: \quad \begin{pmatrix} a_1 b_2 - a_2 b_3 \\ a_1 b_3 - a_2 b_1 \\ a_1 b_1 - a_2 b_2 \end{pmatrix}
\end{align}

\medskip

\noindent \textbf{$\mathbf{2} \otimes \mathbf{3}' = \mathbf{3} \oplus \mathbf{3}'$}
\begin{align}
\mathbf{3}&: \quad \begin{pmatrix} a_1 b_2 - a_2 b_3 \\ a_1 b_3 - a_2 b_1 \\ a_1 b_1 - a_2 b_2 \end{pmatrix} \\
\mathbf{3}'&: \quad \begin{pmatrix} a_1 b_2 + a_2 b_3 \\ a_1 b_3 + a_2 b_1 \\ a_1 b_1 + a_2 b_2 \end{pmatrix}
\end{align}

\medskip

\noindent \textbf{$\mathbf{3} \otimes \mathbf{3} = \mathbf{1} \oplus \mathbf{2} \oplus \mathbf{3} \oplus \mathbf{3}'$}
\begin{align}
\mathbf{1}&: \quad a_1 b_1 + a_2 b_3 + a_3 b_2 \\
\mathbf{2}&: \quad \begin{pmatrix} a_2 b_2 + a_1 b_3 + a_3 b_1 \\ a_3 b_3 + a_1 b_2 + a_2 b_1 \end{pmatrix} \\
\mathbf{3}&: \quad \begin{pmatrix} 2a_1 b_1 - a_2 b_3 - a_3 b_2 \\ 2a_3 b_3 - a_1 b_2 - a_2 b_1 \\ 2a_2 b_2 - a_1 b_3 - a_3 b_1 \end{pmatrix} \\
\mathbf{3}'&: \quad \begin{pmatrix} a_2 b_3 - a_3 b_2 \\ a_1 b_2 - a_2 b_1 \\ a_3 b_1 - a_1 b_3 \end{pmatrix}
\end{align}

\medskip

\noindent \textbf{$\mathbf{3}' \otimes \mathbf{3}' = \mathbf{1} \oplus \mathbf{2} \oplus \mathbf{3} \oplus \mathbf{3}'$}
\begin{align}
\mathbf{1}&: \quad a_1 b_1 + a_2 b_3 + a_3 b_2 \\
\mathbf{2}&: \quad \begin{pmatrix} a_2 b_2 + a_1 b_3 + a_3 b_1 \\ a_3 b_3 + a_1 b_2 + a_2 b_1 \end{pmatrix} \\
\mathbf{3}&: \quad \begin{pmatrix} a_2 b_3 - a_3 b_2 \\ a_1 b_2 - a_2 b_1 \\ a_3 b_1 - a_1 b_3 \end{pmatrix} \\
\mathbf{3}'&: \quad \begin{pmatrix} 2a_1 b_1 - a_2 b_3 - a_3 b_2 \\ 2a_3 b_3 - a_1 b_2 - a_2 b_1 \\ 2a_2 b_2 - a_1 b_3 - a_3 b_1 \end{pmatrix}
\end{align}

\medskip

\noindent \textbf{$\mathbf{3} \otimes \mathbf{3}' = \mathbf{1}' \oplus \mathbf{2} \oplus \mathbf{3} \oplus \mathbf{3}'$}
\begin{align}
\mathbf{1}'&: \quad a_1 b_1 + a_2 b_3 + a_3 b_2 \\
\mathbf{2}&: \quad \begin{pmatrix} a_2 b_2 + a_1 b_3 + a_3 b_1 \\ -a_3 b_3 - a_1 b_2 - a_2 b_1 \end{pmatrix} \\
\mathbf{3}&: \quad \begin{pmatrix} a_2 b_3 - a_3 b_2 \\ a_1 b_2 - a_2 b_1 \\ a_3 b_1 - a_1 b_3 \end{pmatrix} \\
\mathbf{3}'&: \quad \begin{pmatrix} 2a_1 b_1 - a_2 b_3 - a_3 b_2 \\ 2a_3 b_3 - a_1 b_2 - a_2 b_1 \\ 2a_2 b_2 - a_1 b_3 - a_3 b_1 \end{pmatrix}
\end{align}

%%%%%%%%%%%%%%%%%%%%

\section{Construction of Multiplets via Tensor Products}
\label{app:modularforms}

For completeness, in this Appendix we briefly collect the modular multiplets that enter the charged-lepton and neutrino Yukawa sectors. We denote by $Y_a(\tau)$, with $a=1,\dots,5$, the independent weight-2 modular building blocks in the basis adopted in this work. Using the $S_4$ tensor product rules given in Appendix~\ref{S4}, the relevant higher-weight multiplets can be constructed as follows.

At modular weight $k=4$, the multiplets used in the charged-lepton and neutrino sectors are
\begin{equation}
\begin{aligned}
Y_{\mathbf{1}}^{(4)} &= Y_1 Y_2,\\
Y_{\mathbf{2}}^{(4)} &= \left( Y_2^2,\; Y_1^2 \right)^T,\\
Y_{\mathbf{3}}^{(4)} &= \left(
Y_1Y_4-Y_2Y_5,\;
Y_1Y_5-Y_2Y_3,\;
Y_1Y_3-Y_2Y_4
\right)^T,\\
Y_{\mathbf{3}'}^{(4)} &= \left(
Y_1Y_4+Y_2Y_5,\;
Y_1Y_5+Y_2Y_3,\;
Y_1Y_3+Y_2Y_4
\right)^T.
\end{aligned}
\end{equation}

At modular weight $k=6$, a convenient basis for the multiplets appearing in the Majorana sector is
\begin{equation}
\begin{aligned}
Y_{\mathbf{1}}^{(6)} &= 
\left( Y_{\mathbf{2}}^{(2)} \otimes Y_{\mathbf{2}}^{(4)} \right)_{\mathbf{1}}
= Y_1^3+Y_2^3,\\
Y_{\mathbf{2}}^{(6)} &= 
\left( Y_{\mathbf{2}}^{(2)} \otimes Y_{\mathbf{2}}^{(4)} \right)_{\mathbf{2}}
= \left( Y_1^2Y_2,\; Y_1Y_2^2 \right)^T,\\
Y_{\mathbf{3}}^{(6)} &= 
\left( Y_{\mathbf{2}}^{(2)} \otimes Y_{\mathbf{3}'}^{(4)} \right)_{\mathbf{3}}
= \left(
Y_1^2Y_5-Y_2^2Y_4,\;
Y_1^2Y_3-Y_2^2Y_5,\;
Y_1^2Y_4-Y_2^2Y_3
\right)^T,\\
Y_{\mathbf{3}'}^{(6)} &=
\left( Y_{\mathbf{2}}^{(2)} \otimes Y_{\mathbf{3}}^{(4)} \right)_{\mathbf{3}'}
= \left(
Y_1^2Y_5-2Y_1Y_2Y_3+Y_2^2Y_4,\;
Y_1^2Y_3-2Y_1Y_2Y_4+Y_2^2Y_5,\;
Y_1^2Y_4-2Y_1Y_2Y_5+Y_2^2Y_3
\right)^T.
\end{aligned}
\end{equation}

The explicit $\tau$-dependent expressions of the weight-2 building blocks $Y_a(\tau)$ and the complete modular-form basis can be found in Ref.~\cite{Novichkov:2018ovf}.

%%%%%%%%%%%%%%%%%%%%

\section{Details of leptonic Yukawa interactions}

In this Appendix we provide some technical details of the leptonic Yukawa interactions, specifically of how they are expanded after the $S_4$ modular symmetry is spontaneously broken. From the particle content specified in Table \ref{field_assignments}, we can built the following Yukawa terms for the charged lepton sector: 
\begin{eqnarray}
-\mathcal{L}_{Y}^{\left( l\right) } &=&y_{1}^{\left( l\right) }Y_{\mathbf{1}%
}^{\left( 4\right) }\left( \tau \right) \left( \overline{l}_{L}\phi
e_{R}\right) _{\mathbf{1}}+y_{2}^{\left( l\right) }\left[ Y_{\mathbf{2},1}^{\left( 4\right) }\left(
\tau \right) \left( \overline{l}_{L}\phi e_{R}\right) _{\mathbf{2},2}+Y_{%
\mathbf{2},2}^{\left( 4\right) }\left( \tau \right) \left( \overline{l}%
_{L}\phi e_{R}\right) _{\mathbf{2},1}\right]\notag \\
&&+y_{3}^{\left( l\right) }\left[ Y_{\mathbf{3},1}^{\left( 4\right) }\left(
\tau \right) \left( \overline{l}_{L}\phi e_{R}\right) _{\mathbf{3},1}+Y_{%
\mathbf{3},2}^{\left( 4\right) }\left( \tau \right) \left( \overline{l}%
_{L}\phi e_{R}\right) _{\mathbf{3},3}+Y_{\mathbf{3},3}^{\left( 4\right)
}\left( \tau \right) \left( \overline{l}_{L}\phi e_{R}\right) _{\mathbf{3},2}%
\right]\notag \\
&&+y_{4}^{\left( l\right) }\left[ Y_{\mathbf{3}^{\prime },1}^{\left(
4\right) }\left( \tau \right) \left( \overline{l}_{L}\phi e_{R}\right) _{%
\mathbf{3}^{\prime },1}+Y_{\mathbf{3}^{\prime },2}^{\left( 4\right) }\left(
\tau \right) \left( \overline{l}_{L}\phi e_{R}\right) _{\mathbf{3}^{\prime
},3}+Y_{\mathbf{3}^{\prime },3}^{\left( 4\right) }\left( \tau \right) \left( 
\overline{l}_{L}\phi e_{R}\right) _{\mathbf{3}^{\prime },2}\right]+H.c.
\end{eqnarray}
Using the $S_4$ tensor product rules given in Appendix \ref{S4}, the above given charged lepton Yukawa interactions can be rewritten as follows:
\begin{eqnarray}
-\mathcal{L}_{Y}^{\left( l\right) } &=&y_{1}^{\left( l\right) }Y_{\mathbf{1}%
}^{\left( 4\right) }\left( \tau \right) \left( \overline{l}_{1L}\phi e_{1R}+%
\overline{l}_{2L}\phi e_{3R}+\overline{l}_{3L}\phi e_{2R}\right)\notag \\
&&+y_{2}^{\left( l\right) }\left[ Y_{\mathbf{2},1}^{\left( 4\right) }\left(
\tau \right) \left( \overline{l}_{3L}\phi e_{3R}+\overline{l}_{1L}\phi
e_{2R}+\overline{l}_{2L}\phi e_{1R}\right) +Y_{\mathbf{2},2}^{\left(
4\right) }\left( \tau \right) \left( \overline{l}_{2L}\phi e_{2R}+\overline{l%
}_{1L}\phi e_{3R}+\overline{l}_{3L}\phi e_{1R}\right) \right]\notag \\
&&+y_{3}^{\left( l\right) }\left[ Y_{\mathbf{3},1}^{\left( 4\right) }\left(
\tau \right) \left( 2\overline{l}_{1L}\phi e_{1R}-\overline{l}_{2L}\phi
e_{3R}-\overline{l}_{3L}\phi e_{2R}\right) +Y_{\mathbf{3},2}^{\left(
4\right) }\left( \tau \right) \left( 2\overline{l}_{2L}\phi e_{2R}-\overline{%
l}_{1L}\phi e_{3R}-\overline{l}_{3L}\phi e_{1R}\right) \right] \\
&&+y_{3}^{\left( l\right) }Y_{\mathbf{3},3}^{\left( 4\right) }\left( \tau
\right) \left( 2\overline{l}_{3L}\phi e_{3R}-\overline{l}_{1L}\phi e_{2R}-%
\overline{l}_{2L}\phi e_{1R}\right)\notag \\
&&+y_{4}^{\left( l\right) }\left[ Y_{\mathbf{3}^{\prime },1}^{\left(
4\right) }\left( \tau \right) \left( \overline{l}_{2L}\phi e_{3R}-\overline{l%
}_{3L}\phi e_{2R}\right) +Y_{\mathbf{3}^{\prime },2}^{\left( 4\right)
}\left( \tau \right) \left( \overline{l}_{3L}\phi e_{1R}-\overline{l}%
_{1L}\phi e_{3R}\right) +Y_{\mathbf{3}^{\prime },3}^{\left( 4\right) }\left(
\tau \right) \left( \overline{l}_{1L}\phi e_{2R}-\overline{l}_{2L}\phi
e_{1R}\right) \right] +\mathrm{H.c.}\notag
\end{eqnarray}

Regarding the neutrino sector, the corresponding Yukawa terms are:
\begin{eqnarray}
-\mathcal{L}_{Y}^{\left( \nu \right) } &=&y_{1}^{\left( \nu \right) }Y_{%
\mathbf{1}}^{\left( 4\right) }\left( \tau \right) \left( \overline{l}%
_{1L}\eta N_{1R}+\overline{l}_{2L}\eta N_{3R}+\overline{l}_{3L}\eta
N_{2R}\right)   \notag \\
&&+y_{2}^{\left( \nu \right) }\left[ Y_{\mathbf{2},1}^{\left( 4\right)
}\left( \tau \right) \left( \overline{l}_{L}\eta N_{R}\right) _{\mathbf{2}%
,2}+Y_{\mathbf{2},2}^{\left( 4\right) }\left( \tau \right) \left( \overline{l%
}_{L}\eta N_{R}\right) _{\mathbf{2},1}\right]\notag  \\
&&+y_{3}^{\left( \nu \right) }\left[ Y_{\mathbf{3},1}^{\left( 4\right)
}\left( \tau \right) \left( \overline{l}_{L}\eta N_{R}\right) _{\mathbf{3}%
,1}+Y_{\mathbf{3},2}^{\left( 4\right) }\left( \tau \right) \left( \overline{l%
}_{L}\eta N_{R}\right) _{\mathbf{3},3}+Y_{\mathbf{3},3}^{\left( 4\right)
}\left( \tau \right) \left( \overline{l}_{L}\eta N_{R}\right) _{\mathbf{3},2}%
\right]\notag \\
&&+y_{4}^{\left( \nu \right) }\left[ Y_{\mathbf{3}^{\prime },1}^{\left(
4\right) }\left( \tau \right) \left( \overline{l}_{L}\eta N_{R}\right) _{%
\mathbf{3}^{\prime },1}+Y_{\mathbf{3}^{\prime },2}^{\left( 4\right) }\left(
\tau \right) \left( \overline{l}_{L}\eta N_{R}\right) _{\mathbf{3}^{\prime
},3}+Y_{\mathbf{3}^{\prime },3}^{\left( 4\right) }\left( \tau \right) \left( 
\overline{l}_{L}\eta N_{R}\right) _{\mathbf{3}^{\prime },2}\right]\notag  \\
&&+y_{1}^{\left( N\right) }Y_{\mathbf{1}}^{\left( 6\right) }\left( \tau
\right) \left( N_{1R}\sigma ^{\ast }\overline{N_{1R}^{C}}+N_{2R}\sigma
^{\ast }\overline{N_{3R}^{C}}+N_{3R}\sigma ^{\ast }\overline{N_{2R}^{C}}%
\right)   \notag \\
&&+y_{2}^{\left( N\right) }\left[ Y_{\mathbf{2},1}^{\left( 6\right) }\left(
\tau \right) \left( N_{R}\overline{N_{R}^{C}}\right) _{\mathbf{2},2}\sigma
^{\ast }+Y_{\mathbf{2},2}^{\left( 6\right) }\left( \tau \right) \left( N_{R}%
\overline{N_{R}^{C}}\right) _{\mathbf{2},1}\sigma ^{\ast }\right]\notag  \\
&&+y_{3}^{\left( N\right) }\left[ Y_{\mathbf{3},1}^{\left( 6\right) }\left(
\tau \right) \left( N_{R}\overline{N_{R}^{C}}\right) _{\mathbf{3},1}\sigma
^{\ast }+Y_{\mathbf{3},2}^{\left( 6\right) }\left( \tau \right) \left( N_{R}%
\overline{N_{R}^{C}}\right) _{\mathbf{3},3}\sigma ^{\ast }+Y_{\mathbf{3}%
,3}^{\left( 6\right) }\left( \tau \right) \left( N_{R}\overline{N_{R}^{C}}%
\right) _{\mathbf{3},2}\sigma ^{\ast }\right]\notag  \\
&&+y_{4}^{\left( N\right) }\left[ Y_{\mathbf{3}^{\prime },1}^{\left(
6\right) }\left( \tau \right) \left( N_{R}\overline{N_{R}^{C}}\right) _{%
\mathbf{3}^{\prime },1}\sigma ^{\ast }+Y_{\mathbf{3}^{\prime },2}^{\left(
6\right) }\left( \tau \right) \left( N_{R}\overline{N_{R}^{C}}\right) _{%
\mathbf{3}^{\prime },3}\sigma ^{\ast }+Y_{\mathbf{3}^{\prime },3}^{\left(
6\right) }\left( \tau \right) \left( N_{R}\overline{N_{R}^{C}}\right) _{%
\mathbf{3}^{\prime },2}\sigma ^{\ast }\right]\notag  \\
&&+x_{\Omega }Y_{\mathbf{3}^{\prime }}^{\left( 6\right) }\left( \tau \right)
N_{R}\varphi \overline{\Omega _{R}^{C}}+y_{\Omega }Y_{\mathbf{1}}^{\left(
4\right) }(\tau)\Omega _{R}\sigma \overline{\Omega _{R}^{C}}+\mathrm{H.c.}
\end{eqnarray}

The use of the $S_4$ tensor product rules allows to rewrite the above given neutrino interactions as follows:
\begin{eqnarray}
-\mathcal{L}_{Y}^{\left( \nu \right) } &=&y_{1}^{\left( \nu \right) }Y_{%
\mathbf{1}}^{\left( 4\right) }\left( \tau \right) \left( \overline{l}%
_{1L}\eta N_{1R}+\overline{l}_{2L}\eta N_{3R}+\overline{l}_{3L}\eta
N_{2R}\right)   \notag \\
&&+y_{2}^{\left( \nu \right) }\left[ Y_{\mathbf{2},1}^{\left( 4\right)
}\left( \tau \right) \left( \overline{l}_{3L}\eta N_{3R}+\overline{l}%
_{1L}\eta N_{2R}+\overline{l}_{2L}\eta N_{1R}\right) +Y_{\mathbf{2}%
,2}^{\left( 4\right) }\left( \tau \right) \left( \overline{l}_{2L}\eta
N_{2R}+\overline{l}_{1L}\eta N_{3R}+\overline{l}_{3L}\eta N_{1R}\right) %
\right]\notag  \\
&&+y_{3}^{\left( \nu \right) }Y_{\mathbf{3},1}^{\left( 4\right) }\left( \tau
\right) \left( 2\overline{l}_{1L}\eta N_{1R}-\overline{l}_{2L}\eta N_{3R}-%
\overline{l}_{3L}\eta N_{2R}\right)\notag  \\
&&+y_{3}^{\left( \nu \right) }Y_{\mathbf{3},2}^{\left( 4\right) }\left( \tau
\right) \left( 2\overline{l}_{2L}\eta N_{2R}-\overline{l}_{1L}\eta N_{3R}-%
\overline{l}_{3L}\eta N_{1R}\right)\notag  \\
&&+y_{3}^{\left( \nu \right) }Y_{\mathbf{3},3}^{\left( 4\right) }\left( \tau
\right) \left( 2\overline{l}_{3L}\eta N_{3R}-\overline{l}_{1L}\eta N_{2R}-%
\overline{l}_{2L}\eta N_{1R}\right)\notag  \\
&&+y_{4}^{\left( \nu \right) }\left[ Y_{\mathbf{3}^{\prime },1}^{\left(
4\right) }\left( \tau \right) \left( \overline{l}_{2L}\eta N_{3R}-\overline{l%
}_{3L}\eta N_{2R}\right) +Y_{\mathbf{3}^{\prime },2}^{\left( 4\right)
}\left( \tau \right) \left( \overline{l}_{3L}\eta N_{1R}-\overline{l}%
_{1L}\eta N_{3R}\right) +Y_{\mathbf{3}^{\prime },3}^{\left( 4\right) }\left(
\tau \right) \left( \overline{l}_{1L}\eta N_{2R}-\overline{l}_{2L}\eta
N_{1R}\right) \right]\notag  \\
&&+y_{1}^{\left( N\right) }Y_{\mathbf{1}}^{\left( 6\right) }\left( \tau
\right) \left( N_{1R}\overline{N_{1R}^{C}}+N_{2R}\overline{N_{3R}^{C}}+N_{3R}%
\overline{N_{2R}^{C}}\right) \sigma ^{\ast }  \notag \\
&&+y_{2}^{\left( N\right) }\left[ Y_{\mathbf{2},1}^{\left( 6\right) }\left(
\tau \right) \left( N_{3R}\overline{N_{3R}^{C}}+N_{1R}\overline{N_{2R}^{C}}%
+N_{2R}\overline{N_{1R}^{C}}\right) \sigma ^{\ast }+Y_{\mathbf{2},2}^{\left(
6\right) }\left( \tau \right) \left( N_{2R}\overline{N_{2R}^{C}}+N_{1R}%
\overline{N_{3R}^{C}}+N_{3R}\overline{N_{1R}^{C}}\right) \sigma ^{\ast }%
\right]\notag  \\
&&+y_{3}^{\left( N\right) }\left[ Y_{\mathbf{3},1}^{\left( 6\right) }\left(
\tau \right) \left( 2N_{1R}\overline{N_{1R}^{C}}-N_{2R}\overline{N_{3R}^{C}}%
-N_{3R}\overline{N_{2R}^{C}}\right) \sigma ^{\ast }+Y_{\mathbf{3},2}^{\left(
6\right) }\left( \tau \right) \left( 2N_{2R}\overline{N_{2R}^{C}}-N_{1R}%
\overline{N_{3R}^{C}}-N_{3R}\overline{N_{1R}^{C}}\right) \sigma ^{\ast }%
\right]\notag  \\
&&+y_{3}^{\left( N\right) }Y_{\mathbf{3},3}^{\left( 6\right) }\left( \tau
\right) \left( 2N_{3R}\overline{N_{3R}^{C}}-N_{1R}\overline{N_{2R}^{C}}%
-N_{2R}\overline{N_{1R}^{C}}\right) \sigma ^{\ast }\notag \\
&&+x_{\Omega }Y_{\mathbf{3}^{\prime }}^{\left( 6\right) }\left( \tau \right)
N_{R}\varphi \overline{\Omega _{R}^{C}}+y_{\Omega }Y_{\mathbf{1}}^{\left(
4\right) } (\tau) \Omega _{R}\sigma \overline{\Omega _{R}^{C}}+\mathrm{H.c.}
\end{eqnarray}

%%%%%%%%%%%%%%%%%%%%

\section{Neutrino mass matrix}
\label{app:neutrinomass}

The interaction matrix between SM lepton $L$, heavy neutrino $N_R$ and inert scalar $\eta$ is
\begin{eqnarray}
Y_{\nu } &=&\text{{\scriptsize $\left( 
\begin{array}{ccc}
y_{1}^{\left( \nu \right) }Y_{\mathbf{1}}^{\left( 4\right) }\left( \tau
\right) +2Y_{\mathbf{3},1}^{\left( 4\right) }\left( \tau \right)
y_{3}^{\left( \nu \right) } & y_{2}^{\left( \nu \right) }Y_{\mathbf{2}%
,1}^{\left( 4\right) }\left( \tau \right) -y_{3}^{\left( \nu \right) }Y_{%
\mathbf{3},3}^{\left( 4\right) }\left( \tau \right) +y_{1}^{\left( \nu
\right) }Y_{\mathbf{3}^{\prime },3}^{\left( 4\right) }\left( \tau \right)  & 
y_{2}^{\left( \nu \right) }Y_{\mathbf{2},2}^{\left( 4\right) }\left( \tau
\right) -y_{3}^{\left( \nu \right) }Y_{\mathbf{3},2}^{\left( 4\right)
}\left( \tau \right) -y_{4}^{\left( \nu \right) }Y_{\mathbf{3}^{\prime
},2}^{\left( 4\right) }\left( \tau \right)  \\ 
y_{2}^{\left( \nu \right) }Y_{\mathbf{2},1}^{\left( 4\right) }\left( \tau
\right) -y_{3}^{\left( \nu \right) }Y_{\mathbf{3},3}^{\left( 4\right)
}\left( \tau \right) -y_{1}^{\left( \nu \right) }Y_{\mathbf{3}^{\prime
},3}^{\left( 4\right) }\left( \tau \right)  & y_{2}^{\left( \nu \right) }Y_{%
\mathbf{2},2}^{\left( 4\right) }\left( \tau \right) +2Y_{\mathbf{3}%
,2}^{\left( 4\right) }\left( \tau \right) y_{3}^{\left( \nu \right) } & 
y_{1}^{\left( \nu \right) }Y_{\mathbf{1}}^{\left( 4\right) }\left( \tau
\right) -y_{3}^{\left( \nu \right) }Y_{\mathbf{3},1}^{\left( 4\right)
}\left( \tau \right) +y_{4}^{\left( \nu \right) }Y_{\mathbf{3}^{\prime
},1}^{\left( 4\right) }\left( \tau \right)  \\ 
y_{2}^{\left( \nu \right) }Y_{\mathbf{2},2}^{\left( 4\right) }\left( \tau
\right) -y_{3}^{\left( \nu \right) }Y_{\mathbf{3},2}^{\left( 4\right)
}\left( \tau \right) +y_{4}^{\left( \nu \right) }Y_{\mathbf{3}^{\prime
},2}^{\left( 4\right) }\left( \tau \right)  & y_{1}^{\left( \nu \right) }Y_{%
\mathbf{1}}^{\left( 4\right) }\left( \tau \right) -y_{3}^{\left( \nu \right)
}Y_{\mathbf{3},1}^{\left( 4\right) }\left( \tau \right) -y_{4}^{\left( \nu
\right) }Y_{\mathbf{3}^{\prime },1}^{\left( 4\right) }\left( \tau \right)  & 
y_{2}^{\left( \nu \right) }Y_{\mathbf{2},1}^{\left( 4\right) }\left( \tau
\right) +2Y_{\mathbf{3},3}^{\left( 4\right) }\left( \tau \right)
y_{3}^{\left( \nu \right) }%
\end{array}%
\right) $}}.\notag 
\end{eqnarray} 

The interaction matrix between heavy Majorana fermions $N_R$ and $\sigma$ is
\begin{eqnarray}
Y_{N} &=&\left( 
\begin{array}{ccc}
y^{(N)}_{1}Y_{\mathbf{1}}^{\left( 6\right) }\left( \tau \right) +2y^{(N)}_{3}Y_{\mathbf{3%
},1}^{\left( 6\right) }\left( \tau \right)  & y^{(N)}_{2}Y_{\mathbf{2},1}^{\left(
6\right) }\left( \tau \right) -y^{(N)}_{3}Y_{\mathbf{3},3}^{\left( 6\right)
}\left( \tau \right)  & y^{(N)}_{2}Y_{\mathbf{2},2}^{\left( 6\right) }\left( \tau
\right) -y^{(N)}_{3}Y_{\mathbf{3},2}^{\left( 6\right) }\left( \tau \right)  \\ 
y^{(N)}_{2}Y_{\mathbf{2},1}^{\left( 6\right) }\left( \tau \right) -y^{(N)}_{3}Y_{\mathbf{%
3},3}^{\left( 6\right) }\left( \tau \right)  & y^{(N)}_{2}Y_{\mathbf{2},2}^{\left(
6\right) }\left( \tau \right) +2y^{(N)}_{3}Y_{\mathbf{3},2}^{\left( 6\right)
}\left( \tau \right)  & y^{(N)}_{1}Y_{\mathbf{1}}^{\left( 6\right) }\left( \tau
\right) -y^{(N)}_{3}Y_{\mathbf{3},1}^{\left( 6\right) }\left( \tau \right)  \\ 
y^{(N)}_{2}Y_{\mathbf{2},2}^{\left( 6\right) }\left( \tau \right) -y^{(N)}_{3}Y_{\mathbf{%
3},2}^{\left( 6\right) }\left( \tau \right)  & y^{(N)}_{1}Y_{\mathbf{1}}^{\left(
6\right) }\left( \tau \right) -y^{(N)}_{3}Y_{\mathbf{3},1}^{\left( 6\right)
}\left( \tau \right)  & y^{(N)}_{2}Y_{\mathbf{2},1}^{\left( 6\right) }\left( \tau
\right) +2y^{(N)}_{3}Y_{\mathbf{3},3}^{\left( 6\right) }\left( \tau \right) 
\end{array}%
\right),
\end{eqnarray}

and the mass matrix for heavy Majorana fermions $N_R$ is $M_N=\frac{v_\sigma}{\sqrt{2}}Y_N$

The box diagram that generate the effective CP-parity violating operator can be seen in the Fig.~\ref{fig:box}. The amplitude have the form
\begin{equation}
\begin{aligned}
i\mathcal{M}_\mathrm{box}
=
\left(\tilde Y_N\right)_{ij}(\tilde x_\Omega)_j(y_\Omega)^2
I\left(m_{N_j}^2,m_{\Omega_R}^2\right),
\end{aligned}
\end{equation}
where the scalar loop integral $I(a,b)$ is defined as
\begin{equation}
\begin{aligned}
I(a,b)
=
\int\frac{d^4q}{(2\pi)^4}
\frac{ab}{(q^2-a)^2(q^2-b)^2}.
\end{aligned}
\end{equation}

Using Feynman parameters,
\begin{equation}
\begin{aligned}
\frac{1}{(q^2-a)^2(q^2-b)^2}
=
6\int_0^1 dx\,
\frac{x(1-x)}
{\left[q^2-\Delta(x)\right]^4},
\qquad
\Delta(x)=xa+(1-x)b,
\end{aligned}
\end{equation}
it can be expressed in the form
\begin{equation}
\begin{aligned}
I(a,b)
=
6ab\int_0^1 dx\,x(1-x)
\int\frac{d^4q}{(2\pi)^4}
\frac{1}{\left[q^2-\Delta(x)\right]^4}.
\end{aligned}
\end{equation}

In four dimensions, the expression is
\begin{equation}
\begin{aligned}
\int\frac{d^4q}{(2\pi)^4}
\frac{1}{(q^2-\Delta)^4}
=
\frac{i}{16\pi^2}\frac{1}{6\Delta^2},
\end{aligned}
\end{equation}
which yields
\begin{equation}
\begin{aligned}
I(a,b)
=
\frac{i}{16\pi^2}\,
ab\int_0^1 dx\,
\frac{x(1-x)}{\left[xa+(1-x)b\right]^2}.
\end{aligned}
\end{equation}

We thus define the dimensionless loop function
\begin{equation}
\begin{aligned}
J(a,b)
\equiv
ab\int_0^1 dx\,
\frac{x(1-x)}{\left[xa+(1-x)b\right]^2},
\end{aligned}
\end{equation}
whose explicit evaluation gives
\begin{equation}
\begin{aligned}
J(a,b)
=
\frac{ab}{(a-b)^3}
\left[
(a+b)\ln\!\left(\frac{a}{b}\right)-2(a-b)
\right],
\end{aligned}
\end{equation}
and that have a degenerate limit
\begin{equation}
\begin{aligned}
J(a,a)=\frac{1}{6}.
\end{aligned}
\end{equation}

The effective coupling can be written as
\begin{equation}
\begin{aligned}
\lambda_{\mathrm{eff}}
=
-\frac{1}{16\pi^2}
\sum_{i,j=1}^{3}
\left(\tilde Y_N\right)_{ij}(\tilde x_\Omega)_j(y_\Omega)^2\,
J(m_{N_j}^2,m_{\Omega_R}^2).
\end{aligned}
\end{equation}

The effective coupling induces a radiative mass splitting $\delta=\lambda_\text{eff}v_\sigma^2/2$ that can be treated as a small perturbation, so the active neutrino mass matrix can be written as 
\begin{equation}
\begin{aligned}
(M_\nu)_{ij}
=\sum_{k=1}^{3}
\frac{\left(\widetilde{Y}_{\nu}\right)_{ik}\left(\widetilde{Y}_{\nu}\right)_{jk}}{16\pi^2}\,m_{N_k}
\Big[&
\sin^2\beta\,
\big(
F(m_{P_1}^2,m_{N_k}^2)-F(m_{S_1}^2,m_{N_k}^2)
\big)\\
&+
\cos^2\beta\,
\big(
F(m_{P_2}^2,m_{N_k}^2)-F(m_{S_2}^2,m_{N_k}^2)
\big)
\Big],
\end{aligned}
\end{equation}
where the shifts in the scalar mass eigenvalues are obtained from Eqs.~\eqref{eq:physinert} and \eqref{eq:mass}
\begin{equation}
\label{eq:radsplit}
\begin{aligned}
m_{P_1}^2-m_{S_1}^2
=
-2\delta\cos^2\beta,
\qquad
m_{P_2}^2-m_{S_2}^2
=
-2\delta\sin^2\beta.
\end{aligned}
\end{equation}
and with the standard loop function
\begin{equation}
\begin{aligned}
F(m^2,M^2)=\frac{m^2\ln(m^2/M^2)}{M^2-m^2}.
\end{aligned}
\end{equation}

In order to replace the CP-even and CP-odd masses in terms of the tree-level masses, we can define
\begin{equation}
\begin{aligned}
\Delta F_1^{(k)}
&=
F(m_{\chi_1}^2-\delta\cos^2\beta,m_{N_k}^2)
-
F(m_{\chi_1}^2+\delta\cos^2\beta,m_{N_k}^2),
\\[4pt]
\Delta F_2^{(k)}
&=
F(m_{\chi_2}^2-\delta\sin^2\beta,m_{N_k}^2)
-
F(m_{\chi_2}^2+\delta\sin^2\beta,m_{N_k}^2).
\end{aligned}
\end{equation}

Expanding the result up to first order in mass splitting
\begin{equation}
\begin{aligned}
F(m^2-\epsilon,M^2)-F(m^2+\epsilon,M^2)
\simeq
-2\epsilon\,
G(m^2,M^2),
\end{aligned}
\end{equation}
with
\begin{equation}
\begin{aligned}
G(m^2,M^2)
=
\frac{\partial F(m^2,M^2)}{\partial m^2}
=
\frac{
M^2\ln(m^2/M^2)+M^2-m^2
}{
(M^2-m^2)^2
}.
\end{aligned}
\end{equation}

Replacing in the neutrino mass matrix we obtain
\begin{equation}
\begin{aligned}
(M_\nu)_{ij}
&=
\sum_{k=1}^{3}
\frac{\left(\widetilde{Y}_{\nu }\right)_{ik}\left(\widetilde{Y}_{\nu}\right)_{jk}}{32\pi^2}\,m_{N_k}
\left[
\sin^2\beta\,\Delta F_1^{(k)}
+
\cos^2\beta\,\Delta F_2^{(k)}
\right]\\
&\simeq
-\frac{\delta}{16\pi^2}
\sin^2\beta\cos^2\beta
\sum_{k=1}^{3}
\left(\widetilde{Y}_{\nu}\right)_{ik}\left(\widetilde{Y}_{\nu}\right)_{jk}M_{N_k}
\left[
G(m_{\chi_1}^2,M_{N_k}^2)
+
G(m_{\chi_2}^2,M_{N_k}^2)
\right].
\end{aligned}
\end{equation}

\nocite{}
\bibliographystyle{utphys}
\bibliography{references}

\end{document}